\documentclass[10pt,conference]{IEEEtran}
\usepackage{cite}
\usepackage{amsmath,amssymb,amsfonts}
\usepackage{algorithmic}
\usepackage{graphicx}
\usepackage{textcomp}
\usepackage{xcolor}
\usepackage[hyphens]{url}
\usepackage{fancyhdr}
\usepackage{hyperref}
\usepackage{multirow}
\usepackage{multicol}
\usepackage{booktabs}

\usepackage[skins]{tcolorbox}

\usepackage{pifont}
\usepackage{enumitem}
\usepackage{subfigure}

\usepackage{listings,xspace,amsthm}
\newcommand{\hpcayear}{2024}

\definecolor{codegreen}{rgb}{0,0.6,0}
\definecolor{codegray}{rgb}{0.5,0.5,0.5}
\definecolor{codepurple}{rgb}{0.58,0,0.82}
\definecolor{backcolour}{rgb}{0.95,0.95,0.92}
\lstdefinestyle{mystyle}{
    backgroundcolor=\color{backcolour},
    basicstyle=\ttfamily\scriptsize,
    breakatwhitespace=false,
    captionpos=b,
    keepspaces=true,
    numbers=left,
    numbersep=2pt,
    mathescape,
    xleftmargin=2em,
    xrightmargin=0em,
    numberstyle=\scriptsize
}
\usepackage{stmaryrd}
\usepackage{comment}

\newcommand{\tool}{{\sc FIRMER}\xspace}

\newcommand{\dom}{\mathbb{B}}
\newcommand{\Gone}{{\tt 1}\xspace}
\newcommand{\Gzero}{{\tt 0}\xspace}
\newcommand{\Gand}{{\tt and}\xspace}
\newcommand{\Gor}{{\tt or}\xspace}
\newcommand{\Gnor}{{\tt nor}\xspace}
\newcommand{\Gnand}{{\tt nand}\xspace}
\newcommand{\Gxor}{{\tt xor}\xspace}
\newcommand{\Gxnor}{{\tt xnor}\xspace}
\newcommand{\Gnot}{{\tt not}\xspace}
\newcommand{\Gnnot}{{\tt nnot}\xspace}
\newcommand{\Gset}{{\tt set}\xspace}
\newcommand{\Greset}{{\tt reset}\xspace}
\newcommand{\Gates}{\mathbb{G}}

\newcommand{\gate}{{\tt g}}

\newcommand{\set}{\sqcap}
\newcommand{\reset}{\sqcup}
\newcommand{\nneg}{\overline{\neg}}
\newcommand{\SeqC}{\mathcal{S}}
\newcommand{\CC}{\mathcal{C}}
\newcommand{\Cc}{{\tt c}}
\newcommand{\Rr}{{\tt r}}
\newcommand{\Nn}{{\tt n}}
\newcommand{\CR}{{\tt cr}}

\newcommand{\VV}{{\sf V}}
\newcommand{\EE}{{\sf e}}
\newcommand{\II}{\mathcal{I}}
\newcommand{\OO}{\mathcal{O}}
\newcommand{\RR}{\mathcal{R}}

\newcommand{\BB}{{\bf B}}

\newcommand{\nwedge}{\overline{\wedge}}
\newcommand{\nvee}{\overline{\vee}}
\newcommand{\noplus}{\overline{\oplus}}

\newcommand{\sem}[1]{\llbracket#1\rrbracket}

\newcommand{\myding}[1]{\textcolor{red}{\ding{#1}}}

\newcommand{\thy}[1]{{\color{blue}\textbf{thy:} #1 \textbf{}\color{black}}}

\newtheorem{definition}{Definition}
\newtheorem{theorem}{Theorem}

\newtheorem{proposition}{Proposition}
\newtheorem{corollary}{Corollary}
\newtheorem{example}{Example}

\newcommand{\hpcasubmissionnumber}{NaN}
\title{SAT-based Formal Fault-Resistance Verification of Cryptographic Circuits}

\def\hpcacameraready{} 

\newcommand\hpcaauthors{Huiyu Tan$\dagger$ and Pengfei Gao$\dagger$ and Taolue Chen$\ddagger$ and Fu Song$\dagger$ and Zhilin Wu$\natural$}
\newcommand\hpcaaffiliation{ShanghaiTech University$\dagger$, Birkbeck, University of London$\ddagger$, \\ State Key Laboratory of Computer Science, Institute of Software, Chinese Academy of Sciences$\natural$}
\newcommand\hpcaemail{songfu@shanghaitech.edu.cn}

\author{
  \ifdefined\hpcacameraready
    \IEEEauthorblockN{\hpcaauthors{}}
      \IEEEauthorblockA{
        \hpcaaffiliation{} \\
        \hpcaemail{}
      }
  \else
    \IEEEauthorblockN{\normalsize{HPCA \hpcayear{} Submission
      \textbf{\#\hpcasubmissionnumber{}}} \\
      \IEEEauthorblockA{
        Confidential Draft \\
        Do NOT Distribute!!
      }
    }
  \fi
}

\fancypagestyle{camerareadyfirstpage}{%
  \fancyhead{}
  
  \fancyfoot[C]{}
}
\begin{document}
\maketitle

\ifdefined\hpcacameraready
  \thispagestyle{camerareadyfirstpage}
  \pagestyle{empty}
\else
  \thispagestyle{plain}
  \pagestyle{plain}
\fi

\newcommand{\hpcaheight}{0mm}
\ifdefined\eaopen
\renewcommand{\hpcaheight}{12mm}
\fi


\begin{abstract}

Fault injection attacks represent a type of active, physical attack against cryptographic circuits.
Various countermeasures have been proposed to thwart such attacks, the design and implementation of which are, however, intricate, error-prone, and laborious. 
The current formal fault-resistance verification approaches are limited in efficiency and scalability. In this paper,
we formalize the fault-resistance verification problem which is shown to be NP-complete. 
We then devise novel approach for encoding the fault-resistance verification problem as the Boolean satisfiability (SAT) problem so that 
off-the-shelf SAT solvers can be utilized. The approach is implemented in an open-source tool \tool
which is  evaluated extensively on realistic cryptographic circuit benchmarks.
The experimental results show that \tool is able to verify fault-resistance of almost all (46/48) benchmarks in 3 minutes (the other two is verified in 35 minutes). 
In contrast, the prior approach fails on 23 fault-resistance verification tasks even after 24 hours (per task).

\end{abstract}

\section{Introduction}
\label{sec:intro}
Cryptographic circuits have been  
widely used in providing secure authentication, privacy, and integrity,
due to rising security risks in sensor networks, healthcare, cyber-physical systems, and the Internet of Things~\cite{AtzoriIM10,TYAGI202122,nist}.
However, cryptographic circuits are vulnerable to various effective physical attacks, and remains an open challenge even after two decades of research.
This paper focuses on an infamously effective attack, i.e., fault injection attacks~\cite{biham1997differential,BDFGZ14,BHL17,Baksi2022}.

Fault injection attacks deliberately inject disturbances into a cryptographic circuit when it is running cryptographic computation, and analyze the information from the correct (non-faulty) and the incorrect (faulty) outputs, attempting to deduce information on the secret key. Fault injection attacks allow the adversary to bypass certain assumptions 
in classical cryptanalysis methods
where the cipher is considered to be a black box and therefore cannot be tampered with.
The disturbances could be injected in various different ways, such as clock glitches~\cite{AgoyanDNRT10,EndoSHAS11,SelmkeHO19}, underpowering~\cite{elmaneGD08},
voltage glitches~\cite{zussa2013power}, electromagnetic pulses~\cite{DehbaouiDRT12,DumontLM19,DumontLM21}
and laser beams~\cite{skorobogatov2003optical,RoscianSDT13,CourbonLFT14,SchellenbergFGH16,DutertreBCCFFGH18}.
Secret information can be deduced by differential fault analysis~\cite{biham1997differential},
ineffective fault analysis~\cite{clavier2007secret}, statistical fault analysis~\cite{fuhr2013fault}, and statistical ineffective fault analysis~\cite{dobraunig2018statistical}.
Thus, fault injection attacks pose a severe security threat to embedded computing devices with cryptographic modules.

Both detection-based and correction-based countermeasures have been proposed to defend against fault injection attacks~\cite{Malkin2006ACC,Aghaie2020ImpeccableC,Shahmirzadi2020ImpeccableCI}.
The former aims to detect fault injections and infect the output result in the presence of faults in order to make it nonexploitable by an attacker with an error flag output; 
the latter aims to correct the faulty cryptographic computation in the presence of faults.
An effective countermeasure must be \emph{fault-resistance}, i.e., detecting or correcting faults in time once some faults occur. Designing and implementing secure, efficient, and low-cost cryptographic circuits
is notably error-prone, hence it is crucial to rigorously verify
fault-resistance, especially  
at the gate level (which is closer to the final circuit sent to the fab for the tape-out).
Typically this is done at the last stage of the front-end design so the flaws introduced by front-end tools (e.g., optimization passes) can be detected.

There is more specialized work for assuring fault-resistance, (e.g.,~\cite{SimevskiKK13,BurchardGEH00KP17,saha2018expfault,arribas2020cryptographic,khanna2017xfc,srivastava2020solomon,wang2021sofi,nasahl2022synfi}), but almost all of them
focus on finding flaws or checking the effectiveness of user-specified fault test vectors (i.e., certain instances).
In principle, to achieve completeness, all the possible test vectors (varying in fault types, injected gates and clock cycles) 
must be 
checked under all valid input combinations, which is virtually infeasible in practice.
To alleviate this issue, recently, a binary decision diagram (BDD)~\cite{Bryant86} based approach, called FIVER~\cite{RichterBrockmann2021FIVERR} was proposed, 
which does not explicitly enumerate all valid input combinations and is optimized to avoid some fault test vectors.
However, it still has to repeatedly build BDD models for a huge number of fault test vectors, failing to verify relatively larger circuits in a reasonable amount of time.
(For instance, it fails to prove fault-resistance of a single-round 2-bit protected AES in 24 hours.)
 
\noindent
{\bf Contributions.} In this work, 
we first formalize the fault-resistance verification problem, 
which lays a solid foundation for the subsequent reasoning. 
We show that this problem is NP-complete. 
Highlighted by this computational complexity, 
we then propose a novel SAT-based approach for verifying fault-resistance
by utilizing modern efficient SAT solvers.
The encoding for fault-resistance verification into SAT solving requires a novel treatment.
Technically, with
a countermeasure and a fault-resistance model, we generate a new conditionally-controlled faulty circuit, which is in turn reduced
to the SAT problem. 
Intuitively we replace each vulnerable gate with a designated gadget (i.e., sub-circuit) with (1) a control input for controlling if a fault is injected on the gate,
and (2) selection inputs for choosing which fault type is injected. This approaches
avoids explicit enumeration of all the possible fault test vectors and can fully utilize
the conflict-driven clause learning (CDCL) feature of modern SAT solvers.
Furthermore, we introduce a 
reduction technique to safely reduce the number of vulnerable gates when verifying fault-resistance, which significantly improves the verification efficiency.

We implement our approach in an open-source tool \tool (\textbf{F}ault \textbf{I}njection counte\textbf{R} \textbf{M}easure verifi\textbf{ER}), which is 
based on Verilog gate-level netlists.
We evaluate \tool on 19 realistic cryptographic circuits (i.e., rounds of AES, CRAFT and LET64) with both detection- and correction-based countermeasures, where 
the number of gates range from 925 to 40,184. The results show
that our approach is effective and efficient in verifying the fault-resistance against various fault-resistant models.
Almost all the benchmarks (46 out of 48) can be verified in less than 3 minutes (except for two which take
31 and 35 minutes respectively). In comparison,
FIVER runs out of time (with timeout 24 hours) on 23 (out of 48) fault-resistance verification tasks.

To summarize, we make the following major contributions:
\begin{itemize}[leftmargin=*]
  \item We formalize the fault-resistance verification problem 
  and identify its  NP-complete computational complexity for the first time. 
  \item We propose a novel SAT-based approach for formally verifying fault-resistance with an accelerating technique. 
  \item We implement an open-source tool for verifying fault-resistance in Verilog gate-level netlists.
  \item We extensively evaluate our tool on realistic cryptographic circuits to show its effectiveness and efficiency.
\end{itemize}

\smallskip
\noindent
{\bf Outline}.
Section~\ref{sec:preliminary} briefly recaps circuits, fault injection attacks and their countermeasure.
Section~\ref{sec:problem} formulates the fault-resistance verification problem and studies its complexity. 
Section~\ref{sec:SAT-based} presents our SAT-based verification approach.
Section~\ref{sec:experiments} reports experimental results.
We discuss related work in Section~\ref{chap:related work} and finally conclude this work in Section~\ref{chap:conclusion}.

\section{Preliminary}
\label{sec:preliminary}

In this section, we introduce (gate-level) circuits, fault injection attacks
and their countermeasures.

\subsection{Notations}
We denote by $\dom$ the Boolean domain $\{\Gzero,\Gone\}$
and by $[n]$ the set of integers $\{1,\cdots, n\}$ for any integer $n\geq 1$.
To describe  standard circuits, we consider the logic gates:
\Gand ($\wedge$), \Gor ($\vee$), \Gnand ($\nwedge$), \Gnor ($\nvee$), \Gxor ($\oplus$), \Gxnor ($\noplus$), and \Gnot ($\neg$),
all of which are binary gates except for \Gnot.
We note that $\overline{\bullet}(x_1,x_2)=\neg \bullet(x_1,x_2)$,
so $\overline{\overline{\bullet}}$ may be used to denote $\bullet$ for any $\bullet\in\{\wedge,\vee,\oplus\}$.
The set of these gates is denoted by $\Gates$.
In addition, we introduce three auxiliary logic gates to describe faulty circuits:
\Gnnot ($\nneg$), \Gset ($\set$) and \Greset ($\reset$),
where $\nneg x=x$, $\set$ and $\reset$ are two constant logic gates whose outputs are $\Gone$ and $\Gzero$, respectively.
The extended set of logic gates is denoted by $\Gates^\star$. 

\subsection{Synchronous Circuits}
\label{circuitmodel}
 
\begin{definition}[Combinational circuit]
A \emph{combinational circuit} $C$ is a tuple
$(V,I,O, E,\gate)$, where
\begin{itemize}
  \item $V$ is a set of vertices, $I\subset V$ is a  set of inputs, and $O\subset V$ is a set of outputs; 
  \item $E\subseteq (V\setminus O) \times (V\setminus I)$ is a set of edges
each of which represents a wire connecting two vertices and carrying a digital signal
from the domain $\dom$;
\item $(V,E)$ forms a directed acyclic graph (DAG);
  \item each internal vertex $v\in V\setminus (I\cup O)$ is a logic gate associated with its function $\gate(v)\in \Gates^\star$ whose fan-in size
  is equal to the in-degree of the vertex $v$.
\end{itemize}
\end{definition}

Intuitively, combinational circuits represent Boolean
functions. 
The behavior of a combinational circuit is
memoryless, namely, the outputs depend solely on the
 inputs and are independent of the circuit's past history.
The semantics of the combinational circuit $C$ is described by the associated
Boolean function $\sem{C}:\dom^{|I|}\rightarrow \dom^{|O|}$
such that for any signals  $\vec{x}\in \dom^{|I|}$ of the inputs $I$,
$\sem{C}(\vec{x})=\vec{y}$ iff under the input signals $\vec{x}$ the output signals $O$ of the circuit $C$ are $\vec{y}$ .
 
A (synchronous) sequential circuit is a combinational circuit with feedback synchronized by a global clock.
It has primary inputs, primary outputs,
a combinational circuit and memory in the form of registers (or flip-flops).
The output signals of registers at a clock cycle represent an internal state.
At each clock cycle, the combinational circuit produces its output using the current internal state
and the primary inputs as its inputs. The output comprises two parts: one is used as primary outputs
while the other is stored in the registers, which will be the internal state for the next clock cycle and
can be seen as the feedback of the combinational circuit to the next clock cycle.

We focus on round-based circuit implementations of cryptographic primitives
so that the synchronous circuits always have bounded clock cycles
and can be unrolled  
by clock cycles.  

\begin{definition}[Synchronous circuit]
A \emph{$k$-clock cycle synchronous (sequential) circuit} $\SeqC$ for $k\geq 1$ is a tuple
$(\II,\OO,\RR, \vec{s}_0, \CC)$, where
\begin{itemize}[leftmargin=*]
  \item $\II$ (resp. $\OO$) is a  set of primary inputs (resp. primary outputs);
  \item $\RR=R_0\uplus \cdots \uplus R_k$ is a set of registers, called \emph{memory gates};  
  \item $\vec{s}_0\in \dom^{|R_0|}$ gives initial signals to the memory gates in $R_0$;
  \item $\CC=\{C_1,\cdots, C_k\}$ where $C_i=(V_i,I_i, O_i, E_i,\gate_i)$  for each $i\in[k]$. Moreover,
  the inputs $I_i$ consist of the primary inputs and memory gates $R_{i-1}$,
  the outputs  $O_i$ 
  consists of the primary outputs and memory gates $R_{i}$, and  $V_i\cap V_j=\emptyset$ for any $j\neq i$.
\end{itemize}
\end{definition}

Since 
memory gates are for synchronization only and are essentially the same as the identity function, for the sake of
presentation, we extend the function $\gate_i$ such that
for every memory gate $r\in R_{i-1}$, $\gate_i(r)=\nneg$.
However, we emphasize that it may be changed if fault injections are considered.

A \emph{state} $\vec{s}$ of the circuit $\SeqC$ comprises the output signals of the memory gates.
At any clock cycle $i\in[k-1]$,
given a state $\vec{s}_{i-1}$ and signals $\vec{x}_{i}$ of the primary inputs $\II$, the next state
$\vec{s}_i$
is $\sem{C_i}
(\vec{s}_{i-1},\vec{x})$ projected onto the registers $R_i$ and 
 $\sem{C_i}(\vec{s}_{i-1},\vec{x})$ projected onto $\OO$ gives the primary outputs $\vec{y}_i$.
 In general, we write  $\vec{s}_{i-1}\stackrel{\vec{x}_i|\vec{y}_i}{\longrightarrow}\vec{s}_{i}$ for the state transition at the $i$-th clock cycle.

A \emph{run} $\pi$ under a given sequence of primary inputs $(\vec{x}_{1},\cdots,\vec{x}_{k})$ is a sequence
\[\vec{s}_{0}\stackrel{\vec{x}_1|\vec{y}_1}{\longrightarrow}\vec{s}_{1}\stackrel{\vec{x}_2|\vec{y}_2}{\longrightarrow}\vec{s}_{2}\stackrel{\vec{x}_3|\vec{y}_3}{\longrightarrow}\vec{s}_{3}
\cdots  \vec{s}_{k-1} \stackrel{\vec{x}_k|\vec{y}_k}{\longrightarrow}\vec{s}_{k}.\]
 The \emph{semantics} of the circuit $\SeqC$
  is described by its associated Boolean function
  $\sem{\SeqC}:(\dom^{|\II|})^k\rightarrow (\dom^{|\OO|})^k$
  such that for any sequence of input signals $\vec{x}_1,\cdots, \vec{x}_k\in (\dom^{|\II|})^k$,
 $\sem{\SeqC}(\vec{x}_1,\cdots, \vec{x}_k)=(\vec{y}_1,\cdots, \vec{y}_k)$
 iff 
 \begin{center}
 $\vec{s}_{0}\stackrel{\vec{x}_1|\vec{y}_1}{\longrightarrow}\vec{s}_{1}\stackrel{\vec{x}_2|\vec{y}_2}{\longrightarrow}\vec{s}_{2}\stackrel{\vec{x}_3|\vec{y}_3}{\longrightarrow}\vec{s}_{3}
\cdots  \vec{s}_{k-1} \stackrel{\vec{x}_k|\vec{y}_k}{\longrightarrow}\vec{s}_{k}$.
 \end{center}

We remark that in practice, the combinational circuits $C_i$'s in round-based hardware implementations of a cryptographic primitive
are often similar (many of them are actually the same up to renaming of the vertices),
because the internal rounds of a cryptographic primitive often perform similar computations.
Furthermore, only partial signals of primary inputs $\II$ may be used in one clock cycle and the signals of primary outputs $\OO$ may only be produced in the last clock cycle.
Our formalization ties to be general.
\subsection{Fault Injection Attacks}
\label{fault injection attack}
Fault injection attacks are a type of physical attacks that actively inject faults on some logic and/or memory gates during the execution of a cryptographic circuit
and then statistically analyze the faulty primary outputs to deduce sensitive data such as the
cryptographic key~\cite{Baksi2022}. Over the last two decades, various fault injection mechanisms have been proposed 
such as clock glitches~\cite{AgoyanDNRT10,EndoSHAS11,SelmkeHO19}, underpowering~\cite{elmaneGD08},
voltage glitches~\cite{zussa2013power}, electromagnetic pulses~\cite{DehbaouiDRT12,DumontLM19,DumontLM21}
and laser beams~\cite{skorobogatov2003optical,RoscianSDT13,CourbonLFT14,SchellenbergFGH16,DutertreBCCFFGH18}.

Clock glitch 
causes transient faults in circuits by tampering with a clock signal with
glitches. Under the normal clock, the clock cycle is larger than the maximum path delay in
combinational circuits, allowing full propagation of the signals
so that the input signals to memory gates are stable before the next clock signal triggers the
sampling process of the memory gates. In contrast, under a clock with glitches,
some clock periods are shorter than the maximum path delay 
so the input signals to memory gates become unstable (i.e., only parts of input signals have reached).  
As a result, the memory gates may sample faulty results.

Underpowering and voltage glitches are similar to clock glitches except that
underpowering lowers the supply voltage of the device throughout the entire
execution while voltage glitches only lower the supply voltage for a
limited period of time during the execution.
In contrast to clock glitches that decrease clock periods,
lowering the supply voltage increases the maximum path delay in combinatorial circuits
which can also induce memory gates to sample faulty results.

Electromagnetic pulses induce currents in wire loops
which are the power and ground networks in integrated circuits.
The induced current in a wire loop leads to a (negative or positive) voltage swing between the power
and ground grid. The negative (resp. positive) voltage swings can decrease (resp. increase) the clock signal and the input signals
to memory gates, often leading to reset (resp. set) of the corresponding memory gates, thus injecting faults on memory gates. 
A laser beam on a transistor produces a dense distribution of electron-hole pairs along the laser path,
leading to a reduced voltage and eventually a temporary drift current.
The temporary drift current can be used to alter the output signal of a (logic or memory) gate.

Clock glitches, underpowering and voltage glitches are non-invasive,
as they do not require a modification of the targeted device,
thus are considered as rather inexpensive.
In contrast, electromagnetic pulses and laser beams are semi-invasive, allowing the adversary to inject localized faults,
thus have higher precision than non-invasive attacks, 
but still at reasonable equipment and expertise requirement.

\subsection{Countermeasures against Fault Injection Attacks}\label{sec:countermeasures}
Various countermeasures have been proposed to defend against fault injection attacks.

For clock glitches, underpowering and voltage glitches,
an alternative implementation of the circuit can be developed where signal path delays in combinatorial circuits
are made independent of the sensitive data.
For instance, delay components can be added to
certain signal paths~\cite{GhalatyAS14,EndoLHSOFNKDA15},
or combinational circuits can be reorganized~\cite{EldibWW16}, so that the arrival time of all
output signals of logic gates are independent of the sensitive data.
However, such countermeasures fail to defend against 
electromagnetic pulses and laser beams.

Redundancy-based countermeasures are proposed to detect the presence of a fault.
For instance, spatial redundancy recomputes the output multiple times in parallel~\cite{Malkin2006ACC};
temporal redundancy recomputes the output multiple times consecutively~\cite{Malkin2006ACC},
and information redundancy leverages linear error code from coding theory~\cite{Aghaie2020ImpeccableC}.
Once a fault is detected, the output is omitted or the sensitive data is destroyed, with an error flag signal.
However, such countermeasures are still vulnerable against
advanced fault injection attacks such as Ineffective Fault Attack (IFA)~\cite{Clavier2007SecretEE} and  Statistical Ineffective Fault Analysis (SIFA)~\cite{Dobraunig2018SIFAEI}.
The linear error-code based approach proposed in~\cite{Aghaie2020ImpeccableC}
was extended in \cite{Shahmirzadi2020ImpeccableCI}, which can correct 
faults to protect against IFA and SIFA.

\section{The Fault-Resistance Verification Problem}\label{sec:problem}
We are interested in verifying redundancy based countermeasures.
In this section, we first formalize the fault-resistance verification problem,
and then present an illustrating example.

\subsection{Problem Formulation }
Fix a $k$-clock cycle circuit $\SeqC=(\II,\OO,\RR, \vec{s}_0, \CC)$,
where $\CC=\{C_1,\cdots, C_k\}$ and $C_i=(V_i,I_i, O_i, E_i,\gate_i)$ for each $i\in[k]$.
We assume that $\SeqC$ is a cryptographic circuit  without deploying any countermeasures.
Let $\SeqC'=(\II,\OO',\RR', \vec{s}_0', \CC')$ be the protected counterpart of  $\SeqC$ using 
a redundancy-based countermeasure,
where $\CC'=\{C_1',\cdots, C_k'\}$ and $C_i'=(V_i',I_i', O_i', E_i',\gate_i')$  for each $i\in[k]$. We assume that $\OO'=\OO\cup \{o_{\tt flag}\}$, where $o_{\tt flag}$ is an error flag output indicating whether a fault was detected when
the circuit $\SeqC'$ adopts a detection-based countermeasure. 
If $\SeqC'$ adopts a correction-based countermeasure, i.e., no error flag output is involved,
for clarity, we assume that the error flag output $o_{\tt flag}$ is added but is always $\Gzero$. 

To formalize the fault-resistance verification problem, we first introduce some notations.
We denote by $\BB$ the blacklist of gates that are protected against
fault injection attacks. 
$\BB$ usually contains the gates used in the sub-circuits implementing a detection or correction mechanism.
Note that the effects of faults injected on the other gates can
be propagated into the gates in $\BB$.

To model the effects of different fault injections, we introduce the following three fault types:
\begin{itemize}
  \item bit-set fault  $\tau_{s}$: when injected on a gate, its output signal becomes $\Gone$;
  \item bit-reset fault  $\tau_{r}$: when injected on a gate, its output signal becomes $\Gzero$;
  \item bit-flip fault  $\tau_{bf}$: when injected on a gate, its output signal is flipped, i.e., from $\Gone$ to $\Gzero$ or from $\Gzero$ to $\Gone$.
\end{itemize}

These fault types are able to capture all the effects of faults induced
by both non-invasive fault injections (i.e., clock glitches, underpowering and voltage glitches) and semi-invasive fault injections
(i.e., electromagnetic pulses and laser beams). The detailed discussion refers to~\cite{RichterBrockmann2023RevisitingFA}.
We denote by $\mathcal{T} = \{ \tau_{s},\tau_{r},\tau_{bf}\}$ the set of fault types.

A fault injection with fault type $\tau\in \mathcal{T}$ on a gate
can be exactly characterized by replacing its associated function $\bullet$
with $\tau(\bullet)$:
\[\tau(\bullet):=
\left\{
  \begin{array}{ll}
    \set, & \hbox{if } \tau= \tau_{s};\\
    \reset, & \hbox{if } \tau= \tau_{r};\\
    \overline{\bullet}, & \hbox{if } \tau= \tau_{bf}. \\
  \end{array}
\right.
\]

To specify when, where and how a fault is injected, we introduce fault events.
 \begin{definition}[Fault event]
A \emph{fault event} 
is given by
$\EE(\sigma,\beta,\tau)$, where
\begin{itemize}
  \item $\sigma\in[k]$ specifies the clock cycle of the fault injection, namely, the fault injection occurs at the $\sigma$-th clock cycle;
  \item $\beta\in R_{\sigma-1}'\cup V_\sigma'\setminus (I_\sigma'\cup O_\sigma')$ specifies the gate on which the fault is injected;  we require that $\beta\not\in\BB$;
  \item  $\tau\in \mathcal{T}$ specifies the fault type of the fault injection.
 \end{itemize}
\end{definition}

A fault event $\EE(\sigma,\beta,\tau)$ yields the faulty circuit $$\SeqC'[\EE(\sigma,\beta,\tau)]=(\II,\OO',\RR', \vec{s}_0{'}, \{C_1'',\cdots, C_k''\}),$$ 
where for each $i\in[k]$ and every 
$\beta'\in R_{\sigma-1}'\cup V_\sigma'\setminus (I_\sigma'\cup O_\sigma')$,
\begin{itemize}
  \item $C_i'':=
\left\{
  \begin{array}{ll}
    (V_i',I_i', O_i', E_i'',\gate_i''), & \hbox{if } i= \sigma; \\
    C_i', & \hbox{if } i\neq \sigma;
  \end{array}
\right.$
  \item  $E_i''$ is obtained from $E_i'$ by removing the incoming edges of the gate $\beta$ if $\tau\in\{\tau_s,\tau_r\}$,
  \item $\gate_\sigma''(\beta'):=
\left\{
  \begin{array}{ll}
    \tau(\gate_\sigma'(\beta)), & \hbox{if } \beta'=\beta ; \\
    \gate_\sigma'(\beta), & \hbox{if } \beta'\neq \beta,
  \end{array}
\right.$
\end{itemize}
Intuitively, the faulty circuit $\SeqC'[\EE(\sigma,\beta,\tau)]$ is the same as
the circuit $\SeqC'$ except that the function $\gate_\sigma'(\beta)$ of the gate $\beta$
is transiently replaced by $\tau(\gate_\sigma'(\beta))$ in the $\sigma$-th clock cycle,
while all the other gates at all the clock cycles 
remain the same.
We denote by $\tau(\beta)$ the faulty counterpart of  $\beta$ with fault type $\tau$.

In practice, multiple fault events can occur simultaneously during the same clock cycle
and/or consecutively in different clock cycles, allowing the adversary to conduct sophisticated fault injection attacks.
To formalize this, we introduce fault vectors, as a generalization of
fault events.

\begin{definition}[Fault vector]
A \emph{fault vector} $\VV(\SeqC',\BB,T)$ 
is given by a (non-empty) set of fault events
\begin{center}
$\VV(\SeqC',\BB,T)=\left\{
\begin{array}{c}
  \EE(\alpha_1,\beta_1,\tau_1),\\
  \cdots,\\
  \EE(\alpha_m,\beta_m,\tau_m)
\end{array}
 \mid
 \begin{array}{c}
 \forall i,j\in [m].\\ \beta_i\neq \beta_j\wedge \\\alpha_i\in[k]\wedge \tau_i\in T
 \end{array}
\right\}.$
\end{center}
\end{definition}
A fault can be injected to a gate \emph{at most once}, but multiple faults can be injected to different gates.
in the same or different clock cycles.
Note that $\SeqC'$ is unrolled with clock cycles where each gate is renamed in different clock cycles. Thus, some gates in a fault vector may be the same in the unrolled counterpart.

Similarly, a fault vector $\VV(\SeqC',\BB,T)$ on the circuit $\SeqC'$ yields the faulty circuit
$\SeqC'[\VV(\SeqC',\BB,T)]$, which is obtained by iteratively applying fault events in $\VV(\SeqC',\BB,T)$,
i.e.,
\[\SeqC'[\VV(\SeqC',\BB,T)]:=\SeqC'[\EE(\alpha_1,\beta_1,\tau_1)]\cdots [\EE(\alpha_m,\beta_m,\tau_m)].\]

\begin{definition}[Effectiveness of fault vectors]
A fault vector $\VV(\SeqC',\BB,T)$  is \emph{effective} if there exists a sequence of primary inputs $(\vec{x}_{1},\cdots,\vec{x}_{k})$
such that the sequences of primary outputs $\sem{\SeqC'}(\vec{x}_{1},\cdots,\vec{x}_{k})$
and $\sem{\SeqC'[\VV(\SeqC',\BB,T)]}(\vec{x}_{1},\cdots,\vec{x}_{k})$
differ at some clock cycle which is before the clock cycle when the error flag output $o_{\tt flag}$ differs.
\end{definition}

Intuitively, an effective fault vector breaks the functional equivalence between  $\SeqC$ and $\SeqC'$
and the fault is \emph{not} successfully detected (i.e., setting the error flag output $o_{\tt flag}$).
Note that there are two possible cases for an ineffective fault vector:
either $\sem{\SeqC'}(\vec{x}_{1},\cdots,\vec{x}_{k})$
and $\sem{\SeqC'[\VV(\SeqC',\BB,T)]}(\vec{x}_{1},\cdots,\vec{x}_{k})$
are the same or the fault is successfully detected 
in time.

Hereafter, we denote by $\sharp {\tt Clk}(\VV(\SeqC',\BB,T))$ the size of the set $\{\alpha_1,\cdots,\alpha_m\}$, i.e.,
the number of clock cycles when fault events can occur, and by
${\tt MaxEpC}(\VV(\SeqC',\BB,T))$ the maximum number of fault events per clock cycle,
i.e., $\max_{i\in [m]} |\{\EE(\alpha_i,\beta_i,\tau_i)\in \VV(\SeqC',\BB,T)\}|$. Inspired by the consolidated fault adversary model~\cite{RichterBrockmann2023RevisitingFA},
we introduce the security model of fault-resistance which characterizes
the capabilities of the adversary.

\begin{table*}
\centering 
\caption{Truth table of the S-box in the block cipher RECTANGLE.}\vspace{-1mm}
\label{tab:sbox}
\scalebox{0.9}{\begin{tabular}{|c|cccc cccc cccc cccc|}
  \toprule
  $\vec{x}$                 & 0000 & 0001 & 0010 & 0011 & 0100 & 0101 & 0110 & 0111 & 1000 & 1001 & 1010 & 1011 & 1100 & 1101 & 1110 & 1111 \\ \midrule
  $S(\vec{x})$              & 0110 & 0101 & 1100 & 1010 & 0001 & 1110 & 0111 & 1001 & 1011 & 0000 & 0011 & 1101 & 1000 & 1111 & 0100 & 0010 \\ \midrule
 $S[{\tt s7},\tau_s](\vec{x})$     & \textbf{1}110 & \textbf{1}101 & \textbf{0}100 & \textbf{0}010 & 0001 & \textbf{0}110 & \textbf{1}111 & 1001 & 1011 & \textbf{1}000 & 0011 & \textbf{0}101 & \textbf{0}000 & \textbf{0}111 & \textbf{1}100 & \textbf{1}010 \\
  $S[{\tt s7},\tau_r](\vec{x})$    & 0110 & 0101 & 1100 & 1010 & \textbf{1}001 & 1110 & 0111 & \textbf{0}001 & \textbf{0}011 & 0000 & \textbf{1}011 & 1101 & 1000 & 1111 & 0100 & 0010 \\
  $S[{\tt s7},\tau_{bf}](\vec{x})$ & \textbf{1}110 & \textbf{1}101 & \textbf{0}100 & \textbf{0}010 & \textbf{1}001 & \textbf{0}110 & \textbf{1}111 & \textbf{0}001 & \textbf{0}011 & \textbf{1}000 & \textbf{0}011 & \textbf{0}101 & \textbf{0}000 & \textbf{0}111 & \textbf{1}100 & \textbf{1}010 \\\midrule
   $S[{\tt s9},\tau_s](\vec{x})$    & 0\textbf{0}1\textbf{1} & 0101 & 110\textbf{1} & 101\textbf{1} & 0001 & 111\textbf{1} & 0111 & 1001 & 1011 & 0\textbf{1}0\textbf{1} & 0011 & 1101 & 100\textbf{1} & 1111 & 0\textbf{0}0\textbf{1} & 0\textbf{1}1\textbf{1} \\
  $S[{\tt s9},\tau_r](\vec{x})$     & 0110 & 0\textbf{0}0\textbf{0} & 1100 & 1010 & 000\textbf{0} & 1110 & 0\textbf{0}1\textbf{0} & 1\textbf{1}0\textbf{0} & 1\textbf{1}1\textbf{0} & 0000 & 001\textbf{0} & 110\textbf{0} & \textbf{0}000 & 111\textbf{0} & 0100 & 0010 \\
  $S[{\tt s9},\tau_{bf}](\vec{x})$  & 0\textbf{0}1\textbf{1} & 0\textbf{0}0\textbf{0} & 110\textbf{1} & 101\textbf{1} & 000\textbf{0} & 111\textbf{1} & 0\textbf{0}1\textbf{0} & 1\textbf{1}0\textbf{0} & 1\textbf{1}1\textbf{0} & 0\textbf{1}0\textbf{1} & 001\textbf{0} & 110\textbf{0} & 100\textbf{1} & 111\textbf{0} & 0\textbf{0}0\textbf{1} & 0\textbf{1}1\textbf{1} \\
  \bottomrule
\end{tabular}}\vspace{-4mm}
\end{table*}

\begin{definition}[Fault-resistance model]
A \emph{fault-resistance model} 
is given by $\zeta(\Nn_e,\Nn_c,T,\ell)$, where
\begin{itemize}
  \item $\Nn_e$ is the maximum number of fault events per clock cycle;
  \item $\Nn_c$ is the maximum number of clock cycles in which fault events can occur;
  \item $T\subseteq\mathcal{T}$ specifies the allowed fault types; and
  \item $\ell\in\{\Cc,\Rr,\CR\}$ defines vulnerable gates: $\Cc$ for logic gates in combinational circuits, $\Rr$ for memory gates
  and $\CR$ for both logic and memory gates.
\end{itemize}
\end{definition}

For instance, the fault-resistance model $\zeta(\Nn_e,k,\mathcal{T},\CR)$ gives the strongest capability
to the adversary for a large $\Nn_e$
allowing the adversary to inject faults
to all the gates simultaneously at any clock cycle (except for
those protected in the blacklist $\BB$). The fault-resistance model $\zeta(1,1,\{\tau_{bf}\},\Cc)$
only allows the adversary to choose one logic gate to
inject a bit-flip fault in one chosen clock cycle.
Formally, $\zeta(\Nn_e,\Nn_c,T,\ell)$ defines the set $\sem{\zeta(\Nn_e,\Nn_c,T,\ell)}$ of possible fault vectors
 that can be conducted by the adversary, i.e.,
$\sem{\zeta(\Nn_e,\Nn_c,T,\ell)}$ is
\begin{center}
$\left\{\VV(\SeqC',\BB_\ell,T)\mid
 \begin{array}{c}{\tt MaxEpC}(\VV(\SeqC',\BB_\ell,T))\leq n_e\\
 \wedge \\
  \sharp {\tt Clk}(\VV(\SeqC',\BB_\ell,T))\leq \Nn_c
    \end{array}\right\}$,
\end{center}
where all the fault types involved in the fault vectors are limited in $T$
and \[\BB_\ell:=
\left\{
  \begin{array}{ll}
    \BB, & \hbox{if } \ell=\CR; \\
    \BB\cup \RR, & \hbox{if } \ell=\Cc; \\
    \BB\cup \bigcup_{i\in [k]} V_i'\setminus (I_i'\cup O_i'), & \hbox{if } \ell=\Rr. \\
  \end{array}
\right.
\]

The circuit $\SeqC'$
is \emph{fault-resistant} w.r.t.  a blacklist $\BB$ and a fault-resistance model $\zeta(\Nn_e,\Nn_c,T,\ell)$,
denoted by 
\begin{center}
$\langle\SeqC',\BB\rangle\models \zeta(\Nn_e,\Nn_c,T,\ell),$
\end{center}
if all the fault vectors $\VV(\SeqC',\BB,T)\in \sem{\zeta(\Nn_e,\Nn_c,T,\ell)}$ are ineffective on the circuit $\SeqC'$.

\begin{definition}[Fault-resistance verification problem]
The \emph{fault-resistance verification problem} is to determine if $\langle\SeqC',\BB\rangle\models \zeta(\Nn_e,\Nn_c,T,\ell)$, and in particular,
if $\langle\SeqC',\BB\rangle\models \zeta(\Nn_e,\Nn_c,\mathcal{T},\CR)$.
\end{definition}

\begin{proposition}
If $\langle\SeqC',\BB\rangle\models \zeta(\Nn_e,\Nn_c,\mathcal{T},\CR)$, then $\langle\SeqC',\BB\rangle\models \zeta(\Nn_e,\Nn_c,T,\ell)$
for any $T\subseteq \mathcal{T}$ and any $\ell\in\{\Cc,\Rr,\CR\}$.\qed
\end{proposition}

\begin{theorem}\label{thm:fault-resistance-NP}
The problem of determining whether a circuit $\SeqC'$ is not fault-resistant is
NP-complete.
\end{theorem}

The NP upper bound is relatively easy. 
The NP-hardness is proved by reducing from the SAT problem. For a detailed proof, we refer readers to the appendix. 
 
Note that we focus on \emph{transient} fault events that have a dynamic nature
and become inactive after certain periods or changes in the circuit.
There are \emph{persistent} and \emph{permanent} fault events that have a static nature and will remain active for several or even the entire clock cycles.
As mentioned by~\cite{RichterBrockmann2021FIVERR},
it suffices to consider only transient fault events when modeling fault events,
as a persistent or permanent fault event can be
modeled as a repetitive transient fault event.

\subsection{An Illustrating Example}\label{sec:example}
\begin{figure}[t]
    \centering
    \includegraphics[width=0.95\linewidth]{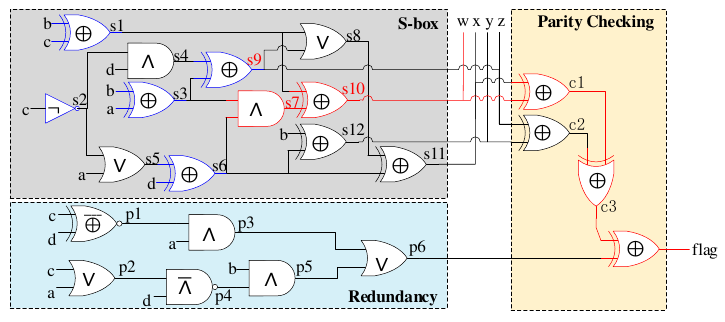}\vspace{-1mm}
    \caption{Circuit representation of the illustrating example.}
    \label{fig:examplecircuit}\vspace{-1mm}
\end{figure}

Consider the S-box used in the cipher RECTANGLE~\cite{zhang2014rectangle}, which is a 4-bit to 4-bit mapping $S:\dom^4\rightarrow \dom^4$ given in Table~\ref{tab:sbox} (the top two rows).
It can be implemented in a combinational circuit as shown in Fig.~\ref{fig:examplecircuit} (grey-area).
It has four 1-bit inputs $\{{\tt a,b,c, d}\}$ denoting the binary representation of the 4-bit input $\vec{x}$,
and four 1-bit outputs $\{{\tt w,x,y,z}\}$ denoting the binary representation of the 4-bit output $S(\vec{x})$, where
${\tt a}$ and ${\tt w}$ are the most significant bits.
The values of the inputs {\tt a,b,c} and {\tt d} depend upon the secret key.

If a fault with fault type $\tau$ is injected on the gate {\tt s7} (i.e., the gate whose output is {\tt s7}),
its function $\gate({\tt s7})$ is changed from $\wedge$ to $\tau(\wedge)$.
As highlighted in red color in Fig.~\ref{fig:examplecircuit},
the effect of this fault will be propagated to the output {\tt w}.
We denote by $S[{\tt s7},\tau]$ the faulty S-box, given in Table~\ref{tab:sbox} for each $\tau\in \mathcal{T}$,
where the faulty output is highlighted in \textbf{bold}.
Since the distribution of the XOR-difference $S[{\tt s7},\tau](\vec{x})\oplus S(\vec{x})$ is biased,
the adversary can narrow down the solutions for $\vec{x}$ according to the value of $S[{\tt s7},\tau](\vec{x})\oplus S(\vec{x})$ which is known to
the adversary.
Finally, the adversary solves  $\vec{x}$ uniquely,
based on which a round key can be obtained (Details refer to~\cite{Baksi2022}).

To thwart single-bit fault injection attacks, one may adopt a single-bit parity protection mechanism~\cite{KarriKG03,BertoniBKMP03,AnaniadisPHBML16}, as shown in
Fig.~\ref{fig:examplecircuit}. The sub-circuit in the blue-area is a redundancy part which computes the Hamming weight of
the output of the S-box from the input but independent on the sub-circuit in the grey-area,
i.e., ${\tt p6}$. 
The sub-circuit in the yellow-area checks the parity of the Hamming weights of $S(\vec{x})$ computed in two independent sub-circuits, i.e.,
${\tt flag}={\tt p6}\oplus {\tt w}\oplus{\tt x}\oplus{\tt y}\oplus{\tt z}$.
If no faults occur, ${\tt flag}$ is $\Gzero$. 

Re-consider the fault injected on the gate {\tt s7}.
We can see that either ${\tt flag}$ becomes $\Gone$, i.e.,
this fault injection can be successfully detected, or the outputs of $S[{\tt s7},\tau](\vec{x})$ and $S(\vec{x})$ are the same,
thus the fault injection is ineffective.

However, the entire circuit is still vulnerable against single-bit fault injection attacks, as
one single-bit fault injection can yield an even number of faulty output bits so that the Hamming weight of the faulty output
remains the same. For instance, the fault injection on the gate {\tt s9} will affects both the outputs {\tt x} and {\tt z}.
As shown in Table~\ref{tab:sbox}, the fault injection cannot be successfully detected if one of the follows holds:
\begin{center}
\small
 \begin{itemize}[leftmargin=*]
  \item $\tau=\tau_s\wedge\vec{x}\in\{0000,1001,1110,1111\}$,
  \item $\tau=\tau_r\wedge\vec{x}\in\{0001,0110,0111,1000,1111\}$,
  \item $\tau=\tau_{bf}\wedge\vec{x}\in\{0000,0001,0110,0111,1000,1001,1110,1111\}$.
\end{itemize}
\end{center}
It is fault-resistant against single-bit fault injection attacks when the blacklist $\BB$ includes
all the logic gates in the parity checking (i.e., yellow-area) and all the logic gates in the S-box (i.e., grey-area) whose out-degree is larger than $2$ (highlighted in blue color in Fig.~\ref{fig:examplecircuit}).
This issue also could be avoided by leveraging the independence property defined by~\cite{Aghaie2020ImpeccableC}, to ensure
an $n$-bit fault injection attack only affect at most $n$ output bits, at the cost of the circuit size. 
The revised implementation is given in the appendix 
which is fault-resistant
against single-bit fault injection attacks when the blacklist $\BB$ includes \emph{only} the logic gates in the parity checking.

\section{SAT-based Fault-resistance Verification}
\label{sec:SAT-based}

In this section, we propose an SAT-based countermeasure verification approach, 
which reduces the fault-resistance verification problems to SAT solving.

\begin{figure}[t]
  \centering
  \includegraphics[width=.48\textwidth]{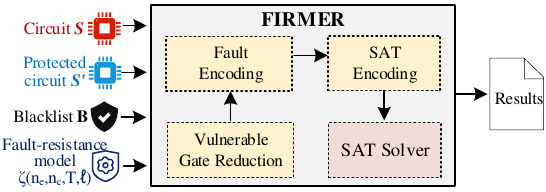}\vspace{-1mm}
  \caption{Framework of our verification approach.}\label{fig:overview}\vspace{-1mm}
\end{figure}

\subsection{Overview}

The overview of our approach is depicted in Fig.~\ref{fig:overview}.  
Given a circuit $\SeqC$ (without any countermeasures), a protected circuit $\SeqC'$ (i.e., $\SeqC$ with a countermeasure), a blacklist $\BB$ of gates on which faults cannot be injected, and a fault-resistance model $\zeta(n_e,n_c,T,\ell)$,
\tool outputs a report on if the protected circuit
$\SeqC'$ is fault-resistant or not.

\tool consists of three key components: vulnerable gate reduction, fault encoding and SAT encoding. 
The vulnerable gate reduction safely reduces the number of vulnerable gates,
thus reduces the size of the resulting Boolean formulas and improves the efficiency.
The fault encoding replaces each vulnerable gate with a gadget (i.e., sub-circuit) with additional
primary inputs controlling if a fault is injected  and selecting a fault type.
The SAT encoding is an extension of the one for checking functional equivalence, where
(i) the maximum number of fault events per clock cycle
and the maximum number of clock cycles in which fault events can occur are both expressed by constraints over control inputs,
and (ii) a constraint on the error flag output is added.
Below,
we present the details of our fault encoding method, 
SAT encoding method 
and vulnerable gate reduction.

 \begin{figure}[t]
   \centering
   \includegraphics[width=0.49\textwidth]{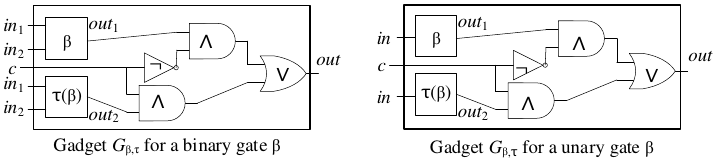}\vspace*{-1mm}
   \caption{Gadgets for encoding fault injections with one fault type.}\label{fig:gadgets}\vspace*{-2mm}
 \end{figure}

\subsection{Fault Encoding}

\noindent{\bf Gadgets}.
To encode a fault injection on a gate $\beta$ with fault type $\tau\in \mathcal{T}$ and $\gate(\beta)=\bullet$,
we define a gadget $G_{\beta,\tau}$ shown in Fig.~\ref{fig:gadgets}. Note that $\tau(\beta)$ denotes the faulty counterpart of the gate $\beta$,
i.e., $\gate(\tau(\beta))=\tau(\bullet)$.
%
Indeed, the gadget $G_{\beta,\tau}$ for a binary gate $\beta$ defines
a Boolean formula $\sem{G_{\beta,\tau}}$ with
\begin{center}
$\sem{G_{\beta,\tau}}(in_1,in_2,c)= \big(c\ ? \ (in_1 \diamond in_2) : (in_1 \bullet in_2) \big),$
\end{center}
where $\diamond=\tau(\bullet)$, and $c$ is a \emph{control input} indicating whether a fault is injected or not.
Namely, $G_{\beta,\tau}$ is equivalent to the faulty gate $\tau(\beta)$ if $c=\Gone$, otherwise $G_{\beta,\tau}$  is equivalent to the original gate $\beta$.
Note that the incoming edges of $\tau(\beta)$ should be omitted if $\tau\in\{\tau_s,\tau_r\}$.
The gadget $G_{\beta,\tau}$ for a unary gate $\beta$ is  defined similarly as
\begin{center}
$\sem{G_{\beta,\tau}}(in,c)= \big(c\ ? \ (\diamond in) : (\bullet in) \big).$
\end{center}

We now generalize the gadget definition to accommodate different fault types $\mathcal{T}=\{\tau_s,\tau_r,\tau_{bf}\}$.
Besides a control input, 
selection inputs are introduced to choose fault types.
The gadget $G_{\beta,\mathcal{T}}$ for a binary logic gate $\beta$ defines
a Boolean formula $\sem{G_{a,\mathcal{T}}}$ such that $\sem{G_{\beta,\mathcal{T}}}(in_1,in_2,c,b_1,b_2)$ is 
\[c\ ? \ \big( b_1\ ?\ (b_2\ ? \ (in_1\diamond in_2) : (in_1 \dagger in_2)) : (in_1 \ddagger in_2) \big) : (in_1 \bullet in_2),\]
where $\diamond=\tau_s(\bullet),\ \dagger=\tau_r(\bullet)$ and $\ddagger=\tau_{bf}(\bullet)$.
Intuitively,
\begin{itemize}
  \item  $c=\Gzero$ means that no fault is injected, i.e., $G_{\beta,\mathcal{T}}$ is equivalent to the original logic gate $\beta$;
  \item  $c=\Gone$ means that a fault is injected. Moreover, the selection inputs $b_1,b_2$ are defined as:
   \begin{itemize}
   \item if $b_1=b_2=\Gone$, then $G_{\beta,\mathcal{T}}$ is equivalent to the faulty logic gate $\tau_s(\beta)$;
   \item if $b_1=\Gone$ and $b_2=\Gzero$, then  $G_{\beta,\mathcal{T}}$ is equivalent to the faulty logic gate $\tau_r(\beta)$;
   \item if $b_1=\Gzero$, then $G_{\beta,\mathcal{T}}$ is equivalent to the faulty logic gate $\tau_{bf}(\beta)$;
  \end{itemize}
\end{itemize}

The gadget $G_{\beta,\mathcal{T}}$ for a unary gate $\beta$  can be defined as
the Boolean formula $\sem{G_{\beta,\mathcal{T}}}$ such that  $\sem{G_{\beta,\mathcal{T}}}(in,c,b_1,b_2)$
is 
\begin{center}
$c\ ? \ \big( b_1\ ?\ (b_2\ ? \ (\diamond in) : (\dagger in)) : (\ddagger in) \big) : (\bullet in).$
\end{center}

For a subset of fault types $T=\{\tau_1,\tau_2\}\subset \mathcal{T}$,
the gadget $G_{\beta,T}$ for a binary or unary gate $\beta$ can be defined accordingly such that
  $\sem{G_{\beta,T}}(in_1,in_2,c,b)$ is
  \begin{center} $c\ ? \ \big( b\ ?\ (in_1 \diamond in_2) : (in_1 \dagger in_2) \big) : (in_1 \bullet in_2)$
  \end{center}
  and 
$\sem{G_{\beta,T}}(in,c,b)$ is 
$c\ ? \ \big( b\ ?\ (\diamond in) : (\dagger in) \big) : (\bullet in)$,
where $\diamond=\tau_1(\bullet)$ and $\dagger=\tau_2(\bullet)$.

We  remark that the faulty counterpart $\tau(\beta)$ of a register $\beta$ is implemented by adding a logic gate so that no additional registers are introduced.
More specifically, $\tau_s(\beta)$ (resp. $\tau_r(\beta)$) is a constant logic gate that always outputs the signal $\Gone$ (resp. $\Gzero$),
and $\tau_{bf}(\beta)$ is a \Gnot logic gate with the incoming edge from the output of the register $\beta$.


\smallskip\noindent
{\bf Conditionally-controlled faulty circuits}.
From $\SeqC'$, 
we construct
a conditionally-controlled faulty circuit $\SeqC''$, where each vulnerable gate
is replaced by a gadget defined above. 

Fix a fault-resistance model $\zeta(n_e,n_c,T,\ell)$.
Assume that the control input $c$ and the set of selection inputs of each gadget $G_{\beta,T}$ are distinct
and different from the ones used in the circuit $\SeqC'$.
We define the conditionally-controlled faulty circuit $\SeqC''$  w.r.t. $\BB$ and $\zeta(n_e,n_c,T,\ell)$
as
$$\SeqC'[\BB,\zeta(n_e,n_c,T,\ell)]:=(\II\uplus\II',\OO',\RR', \vec{s}_0', \CC''),$$
where $\II'=\bigcup_{i\in [k]} I_i''$ 
and $\CC''=\{C_1'',\cdots, C_k''\}$.
For every $i\in [k]$, the circuit $C_i''=(V_i'\uplus V_i'',I_i'\uplus I_i'', O_i', E_i'\uplus E_i'',\gate_i'')$ is obtained from the combinational circuit $C_i'$ as follows.
\begin{tcolorbox}[size=title,colback=white]
   For every gate $\beta\in R_{i-1}'\cup V_i'\setminus (I'_i\cup O'_i)$, if $\beta\not\in \BB_\ell$, then  $\beta$ is replaced by the gadget $G_{\beta,T}$,
    the control and selection inputs of the gadget $G_{\beta,T}$ are added into $I_i''$, the gates and edges of $G_{\beta,T}$ are added into $V_i''$ and $E_i''$ respectively,
    the mapping $\gate_i'$ is expanded to $\gate_i''$ accordingly.
\end{tcolorbox}

Intuitively, a fault vector $\VV(\SeqC',\BB,T)\in \sem{\zeta(n_e,n_c,T,\ell)}$
is encoded as a sequence $(\vec{b}_1,\cdots,\vec{b}_k)$ of the primary inputs $\II'$ for controlling fault types such that $\EE(\alpha,\beta,\tau)\in \VV(\SeqC',\BB,T)$
iff the gadget $G_{\beta,T}$ is equivalent to the faulty gate $\tau(\beta)$ under the inputs $\vec{b}_\alpha$, i.e., the control input of $G_{\beta,T}$ is $\Gone$ and the selection inputs of $G_{\beta,T}$ choose $\tau(\beta)$.
We note that if $\EE(\alpha,\beta,\tau)\not\in \VV(\SeqC',\BB,T)$ for any $\tau\in T$,
then the gadget $G_{\beta,T}$ is equivalent to the original gate $\beta$ under the inputs $\vec{b}_\alpha$, i.e., the control input of the gadget $G_{\beta,T}$ is 
the signal $\Gzero$.

We say that the fault vector $\VV(\SeqC',\BB,T)$ and the sequence $(\vec{b}_1,\cdots,\vec{b}_k)$ of the primary inputs $\II'$ are \emph{compatible}
if the sequence $(\vec{b}_1,\cdots,\vec{b}_k)$ encodes the fault vector $\VV(\SeqC',\BB,T)$.
Note that each sequence $(\vec{b}_1,\cdots,\vec{b}_k)$ of the primary inputs $\II'$
determines a unique compatible fault vector $\VV(\SeqC',\BB,T)$, but
each fault vector $\VV(\SeqC',\BB,T)$ determines a unique compatible sequence $(\vec{b}_1,\cdots,\vec{b}_k)$ of the primary inputs $\II'$
\emph{only} if $T\subset \mathcal{T}$, because  $G_{\beta,\mathcal{T}}$ is equivalent to the faulty logic gate $\tau_{bf}(\beta)$ if $b_1=\Gzero$ no matter the value of $b_2$.
Thus, we can get:

\begin{proposition}\label{prop:encodingfaults}
The number of gates of the circuit $\SeqC''$ (i.e., $\SeqC'[\BB,\zeta(n_e,n_c,T,\ell)]$) is at most $6|T|$ times than that of the circuit $\SeqC'$, and
the following statements hold:
\begin{enumerate}[leftmargin=*]
  \item for each fault vector $\VV(\SeqC',\BB,T)\in \sem{\zeta(n_e,n_c,T,\ell)}$, there exists a compatible sequence $(\vec{b}_1,\cdots,\vec{b}_k)$ of the primary inputs $\II'$
  such that for each sequence $(\vec{x}_{1},\cdots,\vec{x}_{k})$ of primary inputs $\II$,
  \begin{center}\small
  $\sem{\SeqC'[\VV(\SeqC',\BB,T)]}(\vec{x}_{1},\cdots,\vec{x}_{k})=\sem{\SeqC''}((\vec{x}_{1},\vec{b}_1),\cdots,(\vec{x}_{k},\vec{b}_k));$
  \end{center}
  \item for each sequence $(\vec{b}_1,\cdots,\vec{b}_k)$ of the primary inputs $\II'$, there exists a unique compatible fault vector $\VV(\SeqC',\BB,T)\in \sem{\zeta(n_e,n_c,T,\ell)}$
  such that for each sequence $(\vec{x}_{1},\cdots,\vec{x}_{k})$ of primary inputs $\II$,
   \begin{center}\small 
   $\sem{\SeqC'[\VV(\SeqC',\BB,T)]}(\vec{x}_{1},\cdots,\vec{x}_{k})=\sem{\SeqC''}((\vec{x}_{1},\vec{b}_1),\cdots,(\vec{x}_{k},\vec{b}_k)).$
     \end{center}
\end{enumerate}
\end{proposition}

Hereafter, for any sequence $(\vec{b}_1,\cdots,\vec{b}_k)$ of the primary inputs $\II'$, we denote by
$\sharp {\tt Clk}(\vec{b}_1,\cdots,\vec{b}_k)$ the number of clock cycles $i$ such that at least one control input of $\vec{b}_i$ is $\Gone$,
and by ${\tt MaxEpC}(\vec{b}_1,\cdots,\vec{b}_k)$ the maximum sum of the control inputs of $\vec{b}_i$ per clock cycle $i\in[k]$.


\subsection{SAT Encoding}
Recall that 
$\langle\SeqC',\BB\rangle\models \zeta(\Nn_e,\Nn_c,T,\ell)$ iff all the fault vectors $\VV(\SeqC',\BB,T)\in \sem{\zeta(\Nn_e,\Nn_c,T,\ell)}$ are ineffective, 
i.e., for any sequence $(\vec{x}_{1},\cdots,\vec{x}_{k})$ of primary inputs,
either $\sem{\SeqC'}(\vec{x}_{1},\cdots,\vec{x}_{k})=\sem{\SeqC'[\VV(\SeqC',\BB,T)]}(\vec{x}_{1},\cdots,\vec{x}_{k})$
or the fault is successfully detected by setting the error flag output $o_{\tt flag}$ in time.
By Proposition~\ref{prop:encodingfaults}, $\langle\SeqC',\BB\rangle\models \zeta(\Nn_e,\Nn_c,T,\ell)$ iff
for any sequence $((\vec{x}_{1},\vec{b}_1),\cdots,(\vec{x}_{k},\vec{b}_k))$ of primary inputs $\II\cup\II'$
such that 
\begin{itemize}
  \item $\sharp {\tt Clk}(\vec{b}_1,\cdots,\vec{b}_k)\leq \Nn_c$ and ${\tt MaxEpC}(\vec{b}_1,\cdots,\vec{b}_k)\leq \Nn_e$, 
  \item and $\sem{\SeqC'}(\vec{x}_{1},\cdots,\vec{x}_{k})=\sem{\SeqC''}((\vec{x}_{1},\vec{b}_1),\cdots,(\vec{x}_{k},\vec{b}_k))$
or the fault is successfully detected by setting the error flag output $o_{\tt flag}$ in time.
\end{itemize}

The above conditions can be reduced to the SAT problem by adapting the SAT encoding for equivalence checking~\cite{kuehlmann2002combinational,KhasidashviliH03}, with additional constraints
$\sharp {\tt Clk}(\vec{b}_1,\cdots,\vec{b}_k)\leq \Nn_c$ and ${\tt MaxEpC}(\vec{b}_1,\cdots,\vec{b}_k)\leq \Nn_e$.

Formally, the fault-resistance problem of the circuit $\SeqC'$ can be formulated as:
\[\begin{array}{l}
  \forall \vec{x}_1,\cdots,\vec{x}_k\in \dom^{|\II|}.\ \forall \vec{b}_1\in\dom^{|I_1''|},\cdots,\forall\vec{b}_k\in \dom^{|I_k''|}.   \\
\forall i\in[k].\ \forall o\in\OO\setminus \{o_{\tt flag}\}. \\
\quad   \big(\sharp {\tt Clk}(\vec{b}_1,\cdots,\vec{b}_k)\leq \Nn_c\wedge {\tt MaxEpC}(\vec{b}_1,\cdots,\vec{b}_k)\leq \Nn_e\big)\\
\qquad\qquad\qquad\qquad\Rightarrow\big(\psi_{i,o}\neq \psi_{i,o}''\Rightarrow \exists j\in[i].~\psi_{j,o_{\tt flag}}''\big)
\end{array}\]
where 
\begin{itemize}
  \item $\psi_{i,o}$ os a Boolean formula that is satisfiable under the assignment
$(\vec{x}_1,\cdots, \vec{x}_i)$  iff $\sem{\SeqC_i}_{\downarrow o}(\vec{x}_{1},\cdots,\vec{x}_{i})=\Gone$.
  \item $\psi_{i,o}''$ denotes a Boolean formula that is satisfiable under the assignment
$((\vec{x}_1,\vec{b}_1),\cdots, (\vec{x}_i,\vec{b}_i))$ iff $\sem{\SeqC_i''}_{\downarrow o}((\vec{x}_1,\vec{b}_1),\cdots, (\vec{x}_i,\vec{b}_i))=\Gone$.
\end{itemize} 
Intuitively, the above formula is valid iff for any sequence $((\vec{x}_{1},\vec{b}_1),\cdots,(\vec{x}_{k},\vec{b}_k))$ of primary inputs $\II\cup\II'$
such that $\sharp {\tt Clk}(\vec{b}_1,\cdots,\vec{b}_k)\leq \Nn_c$ and ${\tt MaxEpC}(\vec{b}_1,\cdots,\vec{b}_k)\leq \Nn_e$,
if some primary output $o$ (except for the error flag $o_{\tt flag}$) differs at some clock cycle $i$, then the error flag $o_{\tt flag}$ should be $\Gone$
at some clock cycle $j$ with $j\leq i$, i.e., the fault injection is detected in time.

By negating the above formula, the  fault-resistance verification problem is reduced to the satisfiability of the Boolean
formula ($\Psi_{fr}$)
\begin{center}
$\begin{array}{l}
\Psi_{fr}:= \left( \begin{array}{l}\Psi_{\Nn_c}\wedge\Psi_{\Nn_e} \wedge  \bigvee_{i\in[k]}\bigvee_{o\in\OO\setminus \{o_{\tt flag}\}}\\
            \big(\psi_{i,o}\neq \psi_{i,o}'' \wedge \bigwedge_{j\in[i]} \neg\psi_{i,o_{\tt flag}}'' \big),
             \end{array}\right)
 \end{array}$ where
 $\begin{array}{l}
\Psi_{\Nn_c}:=\big(\bigwedge_{i\in[k]} (d_i\Leftrightarrow \bigvee\vec{b}_{i,ctrl}) \big)\wedge \sum_{i\in [k]} d_i\leq \Nn_c,\\
\Psi_{\Nn_e}:=\bigwedge_{i\in[k]} (\sum \vec{b}_{i,ctrl}\leq \Nn_e).
 \end{array}$
\end{center}
and for each $i\in[k]$, $\vec{b}_{i,ctrl}$ denotes the set of control inputs in the primary inputs $\vec{b}_{i}$.
Intuitively, $\Psi_{\Nn_c}$ encodes the constraint $\sharp {\tt Clk}(\vec{b}_1,\cdots,\vec{b}_k)\leq \Nn_c$, where for each $i\in[k]$, $d_i$ is a fresh
Boolean variable such that $d_i$ is $\Gone$ iff some control input in $\vec{b}_{i,ctrl}$ is $\Gone$. Thus, $\sum_{i\in [k]} d_i$
is the total number of clock cycles during which at least one fault is injected on some gate.
$\Psi_{\Nn_e}$ encodes the constraint ${\tt MaxEpC}(\vec{b}_1,\cdots,\vec{b}_k)\leq \Nn_e$, where for each $i\in[k]$, $\sum \vec{b}_{i,ctrl}$ is the total number of faults injected at the $i$-th clock cycle.

Though cardinality constraints of the form $\sum_{i\in[n]} b_i\leq k$ are used in both $\Psi_{\Nn_c}$ and $\Psi_{\Nn_e}$,
they can be efficiently translated into Boolean formulas in polynomial time, and the size of the resulting Boolean formula is also
polynomial in the size of the cardinality constraint~\cite{EenS06,Wynn18}.
In our implementation, we use the sorting network implemented in Z3~\cite{Z3} for translating cardinality constraints into Boolean formulas.


\begin{proposition}
$\langle\SeqC',\BB\rangle\models \zeta(\Nn_e,\Nn_c,T,\ell)$  iff 
the formula $\Psi_{fr}$ is unsatisfiable, where the size of $\Psi_{fr}$ is polynomial in the size of the circuit $\SeqC'$.
\end{proposition}

Note that if the circuit $\SeqC$ is not available, the circuit $\SeqC'$ can be used
for building $\Psi_{fr}$ though the size of $\Psi_{fr}$ may increase. 

\begin{example} \label{emp:SATencoding}
Consider the fault-resistance model $\zeta(1,1,\mathcal{T},\Cc)$. Suppose $S$ is the circuit in Fig.~\ref{fig:examplecircuit} (grey-area),
$S'$ is the entire circuit in Fig.~\ref{fig:examplecircuit}, and the blacklist $\BB$ contains all the logic gates in the redundancy and parity checking parts.
The Boolean formula $\Psi_{fr}$ of the example is:
\begin{center}
$\Psi_{\Nn_c}\wedge \Psi_{\Nn_e}\wedge \big(\bigvee_{o\in\{\tt w,x,y,z\}}\psi_{1,o}\neq \psi_{1,o}''\big) \wedge \neg\psi_{1,{\tt flag}}''\bigwedge_{i=1}^{12} \phi_{i}$
\end{center}
where
\begin{center}\small
\begin{tabular}{ll}
$\Psi_{\Nn_c}:=\big(d_1\Leftrightarrow \bigvee_{i=1}^{12} c_i\big)\wedge d_1\leq 1$,  \quad
$\Psi_{\Nn_e}:= (\sum_{i=1}^{12} c_i\leq 1)$, \\
$\psi_{1,{\tt x}}:= ((b \oplus c) \vee z) \oplus ((\neg c \vee a) \oplus d)$, \\
$\psi_{1,{\tt y}}:= b \oplus (\neg c \vee a) \oplus d$, \qquad\qquad
$\psi_{1,{\tt z}}:=  b \oplus a \oplus (\neg c \wedge d)$, \\
$\psi_{1,{\tt w}}:= (b \oplus c) \oplus ((b \oplus a) \wedge ((\neg c \vee a) \oplus d))$, \\
$\psi_{1,{\tt w}}'':=  g_9$, \quad
$\psi_{1,{\tt y}}'':= g_{10}$, \quad
$\psi_{1,{\tt x}}'':= g_{11}$, \quad
$\psi_{1,{\tt z}}'':= g_{12}$, \\
$\phi_1:=g_1\Leftrightarrow G_{\oplus,\mathcal{T}}''(b,c,c_1,b_{1,1},b_{1,2}),$ \\
$\phi_2:=g_2\Leftrightarrow G_{\oplus,\mathcal{T}}''(b,a,c_2,b_{2,1},b_{2,2}),$ \\
$\phi_3:=g_3\Leftrightarrow G_{\neg,\mathcal{T}}''(c,c_3,b_{3,1},b_{3,2}),$ \\
$\phi_4:=g_4\Leftrightarrow G_{\vee,\mathcal{T}}''(g_3,a,c_4,b_{4,1},b_{4,2}),$ \\
$\phi_5:=g_5\Leftrightarrow G_{\oplus,\mathcal{T}}''(g_4,d,c_5,b_{5,1},b_{5,2}),$  \\
$\phi_6:=g_6\Leftrightarrow G_{\wedge,\mathcal{T}}''(g_2,g_5,c_6,b_{6,1},b_{6,2}),$ \\
$\phi_7:=g_7\Leftrightarrow G_{\vee,\mathcal{T}}''(g_1,z,c_7,b_{7,1},b_{7,2}),$  \\
$\phi_8:=g_8\Leftrightarrow G_{\wedge,\mathcal{T}}''(g_3,d,c_{8},b_{8,1},b_{8,2}),$\\
$\phi_9:=g_9 \Leftrightarrow G_{\oplus,\mathcal{T}}''(g_1,g_6,c_9,b_{9,1},b_{9,2})$,  \\
$\phi_{10}:=g_{10} \Leftrightarrow G_{\oplus,\mathcal{T}}''(b,g_5,c_{10},b_{10,1},b_{10,2})$, \\
$\phi_{11}:=g_{11} \Leftrightarrow G_{\oplus,\mathcal{T}}''(g_7,g_5,c_{11},b_{11,1},b_{11,2})$,  \\
$\phi_{12}:=g_{12} \Leftrightarrow  G_{\oplus,\mathcal{T}}''(g_2,g_8,c_{12},b_{12,1},b_{12,2}),$ \\
 \multicolumn{2}{l}{$\psi_{1,{\tt flag}}'':=\bigoplus_{i=9}^{12} g_i\oplus \Big( \big(a \wedge (c \noplus d)\big) \vee \big(((a \vee c) \nwedge d) \wedge b\big)\Big)$.} \\
\end{tabular}
\end{center}
Note that $g_i$ for each $i\in[12]$ is a fresh Boolean variable as a shortcut of a common gadget via $\phi_i$,
$b_{i,1}$ and $b_{i,2}$ (resp. $c_i$) for each $i\in[12]$ are fresh Boolean variables denoting the selection inputs (resp. control input) of the corresponding gadget,
$\Psi_{\Nn_c}$ can be removed from $\Psi_{fr}$ since it always holds, and
$\Psi_{\Nn_e}$ can be efficiently translated into an equivalent Boolean formula.

We can show that $\Psi_{fr}$ is satisfiable, thus $S'$ is not fault-resistant w.r.t.  $\BB$ and $\zeta(1,1,\mathcal{T},\Cc)$.
Note that in practice, $\BB$ only contains all the logic gates in the parity checking. For the sake of simplicity, $\BB$ also contains all the logic gates in the redundancy part in this example. \qed
\end{example}


\subsection{Vulnerable Gate Reduction}
Consider a fault event $\EE(\alpha,\beta,\tau)$ to the circuit $\SeqC'$.
For any fixed sequence of primary inputs $(\vec{x}_{1},\cdots,\vec{x}_{k})$,
if the output signal of the gate $\beta$ 
does not change, then the fault event $\EE(\alpha,\beta,\tau)$ will not affect
the primary outputs, thus can be omitted.
If it changes, then the effect of $\EE(\alpha,\beta,\tau)$ must be propagated to the successor gates.

Assume the output of the gate $\beta$ is \emph{only} connected to one vulnerable logic gate $\beta'$.
If the output signal of $\beta'$ does not change,
then the effect of the fault event $\EE(\alpha,\beta,\tau)$ is stopped at the gate $\beta'$, thus $\EE(\alpha,\beta,\tau_{bf})$ can be omitted as well.
If 
it changes,
it is flipped either from $\Gone$ to $\Gzero$ or from $\Gzero$ to $\Gone$,  the same effect can be achieved by applying
the fault event $\EE(\alpha,\beta',\tau_{bf})$, or the fault event $\EE(\alpha,\beta',\tau_{s})$ if it is flipped from $\Gzero$ to $\Gone$ or the fault event $\EE(\alpha,\beta',\tau_{r})$ if it is flipped from $\Gone$ to $\Gzero$.
As a result, it suffices to consider fault injections on the gate $\beta'$ instead of both $\beta$ and $\beta'$
when $\tau_{bf}\in T$ or $\{\tau_s,\tau_r\}\subseteq T$,
which reduces the number of vulnerable gates. 


\begin{theorem} \label{thm:faultgatereduce}
Consider a fault-resistance model $\zeta(\Nn_e,\Nn_c,T,\ell)$ such that $\tau_{bf}\in T$ or $\{\tau_s,\tau_r\}\subseteq T$, and $\ell\in\{\Cc,\CR\}$.
Let $\VV_1(\SeqC',\BB,T)=\VV(\SeqC',\BB,T)\cup \{\EE(\alpha,\beta,\tau)\}\in \sem{\zeta(\Nn_e,\Nn_c,T,\ell)}$ be an effective fault vector on the circuit $\SeqC'$.
If the output of the gate $\beta$ is only connected to one logic gate $\beta'\not\in\BB$, 
then 
there exists a fault vector $\VV'(\SeqC',\BB,T)\subseteq \VV(\SeqC',\BB,T)\cup \{\EE(\alpha,\beta',\tau')\}$ for some $\tau'\in T$ such
that $\VV'(\SeqC',\BB,T)$ is also effective on the circuit $\SeqC'$.

Moreover, if $\VV(\SeqC',\BB,T)=\emptyset$, then $\{\EE(\alpha,\beta',\tau')\}$ for some $\tau'\in T$ is effective on the circuit $\SeqC'$.
\end{theorem}

 Theorem~\ref{thm:faultgatereduce} is proved by distinguishing whether the output signal of the gate $\beta'$ differs in the circuits
$\SeqC'$ and $\SeqC'[\VV_1(\SeqC',\BB,T)]$ under the same sequence of primary inputs $(\vec{x}_{1},\cdots,\vec{x}_{k})$.
If it is the same in the circuits
$\SeqC'$ and $\SeqC'[\VV_1(\SeqC',\BB,T)]$ under the same sequence of primary inputs $(\vec{x}_{1},\cdots,\vec{x}_{k})$, then
$\EE(\alpha,\beta,\tau)$ can be removed from $\VV_1(\SeqC',\BB,T)$, otherwise the same effect of $\EE(\alpha,\beta,\tau)$ can be achieved by $\EE(\alpha,\beta',\tau')$ for some $\tau'\in T$.
For a detailed proof, we refer readers to the appendix. 

Let $\BB'$ be the set of gates $\beta$ such that the output of the gate $\beta$ is only connected to one logic gate $\beta'\not\in\BB$,
which can be computed by a graph traversal of the circuit $\SeqC'$.
By applying Theorem~\ref{thm:faultgatereduce}, $\BB'$ can be safely merged with the blacklist $\BB$ while
no protections are required for those gates.


\begin{corollary}\label{cor:faultgatereduce}
Given a fault-resistance model $\zeta(\Nn_e,\Nn_c,T,\ell)$ such that $\tau_{bf}\in T$ or $\{\tau_s,\tau_r\}\subseteq T$, and $\ell\in\{\Cc,\CR\}$,
if $\langle\SeqC',\BB\cup\BB'\rangle\models \zeta(\Nn_e,\Nn_c,T,\ell)$, then $\langle\SeqC',\BB\rangle\models \zeta(\Nn_e,\Nn_c,T,\ell)$.
\end{corollary}


\begin{example} \label{emp:SATencodingwithgatereduce}
Consider the example in Section~\ref{sec:example} and the fault-resistance model $\zeta(1,1,\mathcal{T},\Cc)$.
All the gates in the redundancy part except for
the gate {\tt p6} (i.e., the gate whose output is {\tt p6})
can be added into $\BB'$, as the effect of an effective fault injection on any of those gates
can be achieved by at most one bit-flip fault injection on the gate {\tt p6}. Note that the gate {\tt p6} itself cannot be added into
$\BB'$ because the gate ${\tt flag}$ is in $\BB$.
Similarly, the gates {\tt s4,s5,s7,s8} in the grey-area can be added into $\BB'$,
but the other gates cannot as their outputs are connected to more than one gate or
some outputs of $\{{\tt w,x,y,z}\}$.
Now, the fault-resistance verification problem w.r.t the fault-resistance model $\zeta(1,1,\mathcal{T},\Cc)$
and the blacklist $\BB$ is reduced to SAT solving of
the Boolean formula ($\Psi_{fr}''$)
\begin{center}
$\Psi_{fr}'':=  \Psi_{\Nn_c}'\wedge \Psi_{\Nn_e}'\wedge \big(\bigvee_{o\in\{\tt w,x,y,z\}}\psi_{1,o}\neq \psi_{1,o}''\big) \wedge \neg\psi_{1,{\tt flag}}''  \bigwedge_{i\in Z} \phi_{i}$,
\end{center}
where $Z=\{1,2,3,6,9,10,11,12\}$, $\Psi_{\Nn_c}':=\big(\bigwedge_{i\in Z} (d_i\Leftrightarrow \bigvee\vec{b}_{i,ctrl}) \big)\wedge \sum_{i\in Z} d_i\leq \Nn_c$
and $\Psi_{\Nn_e}':= (\sum_{i\in Z} c_i\leq 1)$.\qed %
\end{example}

\section{Experiments}
\label{sec:experiments}
We have implemented our approach as an open-source tool \tool.
Given circuits $\SeqC$ and $\SeqC'$ in the form of Verilog gate-level netlists,
and a configuration file describing the blacklist and fault-resistance model,
\tool verifies the fault-resistance of $\SeqC'$.
\tool 
first expresses the constraints in quantifier-free bit-vector theory (QF\_Bitvec) using our SAT encoding methods and then
translates to Boolean formulas (in the DIMACS format) via 
Z3~\cite{Z3}.
Those Boolean formulas can be solved by off-the-shelf SAT solvers.
Currently, \tool uses the parallel 
SAT solver, Glucose 4.2.1~\cite{glucose2018}.

  \begin{table}[t]
  \centering\setlength\tabcolsep{1.5pt}
  \caption{Benchmark statistics, where R$i$ denotes $i$ rounds, b$i$ denotes countermeasure against $i$-bit faults, and D (resp. C) denotes detection-based (resp. correction-based) countermeasure.}\vspace{-1mm}
  \label{tab:benchmarks}
\scalebox{0.76}{\begin{tabular}{l|c rrrr|rrrrrrrr} \toprule  
        \textbf{Name} & \textbf{\#Clk} & \textbf{$|\BB|$} & \textbf{\#in} & \textbf{\#out} & \textbf{\#gate} & \textbf{\#and} & \textbf{\#nand} & \textbf{\#or} & \textbf{\#nor} & \textbf{\#xor} & \textbf{\#xnor} & \textbf{\#not} & \textbf{\#reg} \\  \midrule
        AES-R1 & 1 & 0 & 256 & 128 & 21201 & 464 & 7936 & 592 & 8480 & 464 & 560 & 2705 & 0 \\
        AES-R1-b1-D & 1 & 432 & 256 & 129 & 24864 & 576 & 9446 & 560 & 9705 & 828 & 852 & 2897 & 0 \\
        AES-R1-b2-D & 1 & 1055 & 256 & 129 & 34159 & 704 & 12698 & 833 & 13012 & 1440 & 1584 & 3888 & 0 \\\midrule
        CRAFT-R1 & 2 & 0 & 128 & 64 & 480 & 0 & 160 & 16 & 80 & 80 & 32 & 48 & 64 \\
        CRAFT-R1-b1-C & 2 & 192 & 128 & 64 & 3140 & 0 & 864 & 48 & 656 & 460 & 728 & 272 & 112 \\
        CRAFT-R1-b2-C & 2 & 1312 & 128 & 64 & 20948 & 336 & 7856 & 384 & 6576 & 1484 & 2056 & 2080 & 176 \\
        CRAFT-R1-b1-D & 2 & 159 & 128 & 65 & 925 & 41 & 155 & 65 & 148 & 187 & 193 & 56 & 80 \\
        CRAFT-R1-b2-D & 2 & 383 & 128 & 65 & 1522 & 49 & 266 & 49 & 211 & 201 & 539 & 95 & 112 \\
        CRAFT-R1-b3-D & 2 & 511 & 128 & 65 & 1807 & 48 & 282 & 97 & 292 & 240 & 640 & 80 & 128 \\ \midrule
        CRAFT-R2 & 3 & 0 & 192 & 64 & 960 & 0 & 320 & 32 & 160 & 160 & 64 & 96 & 128 \\
        CRAFT-R2-b1-C & 3 & 192 & 192 & 64 & 5848 & 0 & 1680 & 96 & 1248 & 808 & 1264 & 528 & 224 \\
        CRAFT-R2-b2-C & 3 & 1312 & 192 & 64 & 40184 & 672 & 15152 & 752 & 12576 & 2680 & 3936 & 4064 & 352 \\
        CRAFT-R2-b1-D & 3 & 400 & 192 & 65 & 1880 & 100 & 292 & 120 & 311 & 382 & 394 & 121 & 160 \\
        CRAFT-R2-b2-D & 3 & 959 & 192 & 65 & 3139 & 97 & 553 & 100 & 449 & 434 & 1094 & 188 & 224 \\ \midrule
        CRAFT-R3 & 4 & 0 & 256 & 64 & 1440 & 0 & 480 & 48 & 240 & 240 & 96 & 144 & 192 \\
        CRAFT-R3-b3-D & 4 & 1791 & 256 & 65 & 5567 & 192 & 954 & 193 & 836 & 672 & 2048 & 288 & 384 \\ \midrule
        CRAFT-R4 & 5 & 0 & 320 & 64 & 1920 & 0 & 640 & 64 & 320 & 320 & 128 & 192 & 256 \\
        CRAFT-R4-b3-D & 5 & 2303 & 320 & 65 & 7295 & 256 & 1242 & 257 & 1124 & 944 & 2576 & 384 & 512 \\ \midrule
        LED64-R1 & 1 & 0 & 128 & 64 & 976 & 16 & 272 & 16 & 32 & 288 & 304 & 48 & 0 \\
        LED64-R1-b1-D & 1 & 240 & 128 & 65 & 1552 & 16 & 346 & 32 & 53 & 416 & 608 & 81 & 0 \\
        LED64-R1-b2-D & 1 & 575 & 128 & 65 & 2463 & 17 & 479 & 64 & 111 & 512 & 1168 & 112 & 0 \\ \midrule
        LED64-R2 & 2 & 0 & 128 & 64 & 1952 & 32 & 544 & 32 & 64 & 512 & 608 & 96 & 64 \\
        LED64-R2-b1-D & 2 & 480 & 128 & 65 & 2976 & 32 & 690 & 64 & 109 & 760 & 1112 & 129 & 80 \\
        LED64-R2-b2-D & 2 & 1151 & 128 & 65 & 4687 & 33 & 951 & 128 & 231 & 984 & 2072 & 176 & 112 \\ \midrule
        LED64-R3 & 3 & 0 & 128 & 64 & 2928 & 48 & 816 & 48 & 96 & 736 & 912 & 144 & 128 \\
        LED64-R3-b1-D & 3 & 720 & 128 & 65 & 4400 & 48 & 1030 & 96 & 169 & 1116 & 1604 & 177 & 160 \\
        LED64-R3-b2-D & 3 & 1727 & 128 & 65 & 6911 & 49 & 1411 & 192 & 363 & 1492 & 2940 & 240 & 224 \\ \bottomrule
    \end{tabular}}\vspace{-1mm}
\end{table}

We evaluate \tool to answer the following two research questions:
\begin{itemize}
    \item {\bf RQ1.}  How efficient and effective  is \tool for fault-resistance verification?
    \item {\bf RQ2.}  How effective is the vulnerable gate reduction for accelerating fault-resistance verification?
\end{itemize}

\noindent{\bf Benchmarks.}
We use the VHDL
implementations of three cryptographic algorithms (i.e., CRAFT, LED and AES), taken from~\cite{Aghaie2020ImpeccableC,Shahmirzadi2020ImpeccableCI}.
The VHDL implementations are unrolled manually and transformed into Verilog
gate-level netlists using the Synopsys design compiler (version O-2018.06-SP2).
The blacklists are the same as the ones used in FIVER~\cite{RichterBrockmann2021FIVERR}.

The statistics of the benchmark  is given in Table~\ref{tab:benchmarks}, where the first three columns
respectively give the benchmark name, number of clock cycles,
size of the blacklist $\BB$, and the other columns give the numbers of primary inputs, primary outputs, gates and each specific gate.
The CRAFT benchmarks adopts both detection- (D) and correction-based (C) countermeasures.
The CRAFT and LED benchmarks vary with the number of rounds (R$i$) and maximal number of protected faulty bits (b$i$) (i.e., the circuit $\SeqC'$ is claimed to be fault-resistant).
The number of gates ranges from 925 to 40,184 so that the scalability of \tool can be 
evaluated.

The experiments were conducted on a machine with Intel Xeon Gold 6342 2.80GHz CPU, 1T RAM, and Ubuntu 20.04.1. Each verification task is run with 24 hours timeout.

 \begin{table}[t]
    \centering\setlength\tabcolsep{1pt}
  \caption{Results of fault-resistance verification.}\vspace{-1mm}
  \label{tab:fault-resistancemf}
\scalebox{0.68}{
    \begin{tabular}{l|c|rrrrr|r|r|c|c } \toprule
      \multirow{2}{*}{\bf Name}   &  \multirow{2}{*}{\bf Model} & \multicolumn{5}{c|}{\tool ({\bf SAT})} & \textbf{SMT} & \textbf{FIVER} & \multirow{2}{*}{\bf Result} & {\bf Desired}\\ \cline{3-9}
         & & \textbf{\#Var} & \textbf{\# Clause} & \textbf{2CNF} & \textbf{Solving} & \textbf{Total} & \textbf{Time} & \textbf{Time} &  & {\bf Result} \\ \midrule
         AES-R1-b1-D   & $\zeta(1,1, \mathcal{T},\CR)$ & 57340 & 310569 & 1.97 & 193.27 & 195.24 & 4856.58 & \textbf{59.49} & \ding{51} & \ding{51} \\
        AES-R1-b1-D   & $\zeta(2,1,\mathcal{T},\CR)$ & 69088 & 246108 & 1.96 & 0.61 & \textbf{2.57} & 3650.83 & 34744.40 & \myding{55} & \myding{55} \\
        AES-R1-b2-D   & $\zeta(2,1,\mathcal{T},\CR)$ & 86333 & 329543 & 4.84 & 1845.23 & \textbf{1850.07} & 23542.57 & timeout & \ding{51}& \ding{51} \\
        AES-R1-b2-D   & $\zeta(3,1,\mathcal{T},\CR)$ & 90428 & 354018 & 4.38 & 0.91 & \textbf{5.28} & 3386.38 & timeout & \myding{55}& \myding{55}  \\    \midrule
        CRAFT-R1-b1-C & $\zeta(1,1,\mathcal{T},\CR)$ & 6767 & 33938 & 0.13 & 0.66 & 0.79 & 9.81 & \textbf{0.14} & \ding{51}  & \ding{51}\\
        CRAFT-R1-b1-C & $\zeta(2,1,\mathcal{T},\CR)$ & 7608 & 32039 & 0.14 & 0.06 & \textbf{0.20} & 0.28 & 1.34 & \myding{55} & \myding{55} \\
        CRAFT-R1-b2-C & $\zeta(2,1,\mathcal{T},\CR)$ & 52901 & 234385 & 2.79 & 130.00 & \textbf{132.79} & 2898.55 & 338.23 & \ding{51}& \ding{51} \\
        CRAFT-R1-b2-C & $\zeta(3,1,\mathcal{T},\CR)$ & 55099 & 258959 & 2.80 & 0.59 & \textbf{3.39} & 21.49 & 4491.08 & \myding{55}& \myding{55}  \\
        CRAFT-R1-b1-D & $\zeta(1,1,\mathcal{T},\CR)$ & 2255 & 10735 & 0.03 & 0.10 & 0.13 & 0.33 & \textbf{0.04} & \ding{51} & \ding{51} \\
        CRAFT-R1-b1-D & $\zeta(2,1,\mathcal{T},\CR)$ & 2567 & 10316 & 0.03 & 0.03 & 0.06 & \textbf{0.01} & 0.35 & \myding{55}& \myding{55} \\
        CRAFT-R1-b2-D & $\zeta(2,1,\mathcal{T},\CR)$ & 3356 & 13780 & 0.02 & 0.22 & \textbf{0.24} & 1.02 & 1.13 & \ding{51} & \ding{51}\\
        CRAFT-R1-b2-D & $\zeta(3,1,\mathcal{T},\CR)$ & 3538 & 15058 & 0.03 & 0.03 & \textbf{0.06} & 0.20 & 117.54 & \myding{55} & \myding{55}\\
        CRAFT-R1-b3-D & $\zeta(3,1,\mathcal{T},\CR)$ & 3974 & 19042 & 0.03 & 0.43 & \textbf{0.46} & 4.43 & 8007.42 & \ding{51}  & \ding{51} \\
        CRAFT-R1-b3-D & $\zeta(4,1,\mathcal{T},\CR)$ & 4132 & 20428 & 0.04 & 0.03 & \textbf{0.07} & 0.34 & 82955.17 & \myding{55}  & \myding{55}\\    \midrule
        CRAFT-R2-b1-C & $\zeta(1,1,\mathcal{T},\CR)$ & 12644 & 63451 & 0.36 & 11.27 & 11.63 & 101.19 & \textbf{4.20} & \ding{51}& \ding{51}  \\
        CRAFT-R2-b1-C & $\zeta(2,1,\mathcal{T},\CR)$ & 14215 & 59873 & 0.31 & 0.14 & \textbf{0.45} & 0.60 & 394.79 & \myding{55} & \myding{55}\\
        CRAFT-R2-b2-C & $\zeta(2,1,\mathcal{T},\CR)$ & 104139 & 452464 & 6.68 & 2054.13 & \textbf{2060.81} & 13585.63 & 5012.34 & \ding{51}& \ding{51} \\
        CRAFT-R2-b2-C & $\zeta(3,1,\mathcal{T},\CR)$ & 108384 & 498341 & 7.74 & 2.67 & \textbf{10.41} & 3000.80 & 29120.79 & \myding{55} & \myding{55} \\
        CRAFT-R2-b1-D & $\zeta(1,1,\mathcal{T},\CR)$ & 4195 & 20279 & 0.06 & 0.88 & \textbf{0.94} & 6.16 & 1.04 & \ding{51}& \ding{51}  \\
        CRAFT-R2-b1-D & $\zeta(2,1,\mathcal{T},\CR)$ & 4795 & 19338 & 0.07 & 0.03 & \textbf{0.10} & 0.18 & 23.63 & \myding{55} & \myding{55} \\
        CRAFT-R2-b1-D & $\zeta(1,2,\mathcal{T},\CR)$ & 4195 & 20279 & 0.06 & 1.76 & \textbf{1.82} & 8.25 & 1291.12 & \ding{51} & \ding{51}\\
        CRAFT-R2-b1-D & $\zeta(1,3,\mathcal{T},\CR)$ & 4194 & 20277 & 0.06 & 1.35 & \textbf{1.41} & 7.19 & timeout & \ding{51} & \ding{51}\\
        CRAFT-R2-b2-D & $\zeta(2,1,\mathcal{T},\CR)$ & 5959 & 25370 & 0.09 & 3.25 & \textbf{3.34} & 34.93 & 88.35 & \ding{51}& \ding{51} \\
        CRAFT-R2-b2-D & $\zeta(3,1,\mathcal{T},\CR)$ & 6268 & 27935 & 0.11 & 0.06 & \textbf{0.17} & 1.38 & 9141.86 & \myding{55} & \myding{55}\\
        CRAFT-R2-b2-D & $\zeta(2,2,\mathcal{T},\CR)$ & 5959 & 25370 & 0.11 & 3.47 & \textbf{3.58} & 43.36 & timeout & \ding{51} & \ding{51} \\    \midrule
        CRAFT-R3-b3-D & $\zeta(3,1,\mathcal{T},\CR)$ & 10364 & 48541 & 0.20 & 37.66 & \textbf{37.86} & 414.00 & timeout & \ding{51} & \ding{51}\\
        CRAFT-R3-b3-D & $\zeta(4,1,\mathcal{T},\CR)$ & 10760 & 51937 & 0.21 & 0.10 & \textbf{0.31} & 4.69 & timeout & \myding{55}  & \myding{55} \\
        CRAFT-R3-b3-D & $\zeta(3,2,\mathcal{T},\CR)$ & 10366 & 48560 & 0.20 & 57.54 & \textbf{57.74} & 393.80 & timeout & \ding{51}& \ding{51} \\
        CRAFT-R3-b3-D & $\zeta(3,3,\mathcal{T},\CR)$ & 10364 & 48541 & 0.21 & 36.83 & \textbf{37.04} & 357.59 & timeout & \ding{51} & \ding{51} \\
        CRAFT-R3-b3-D & $\zeta(3,4,\mathcal{T},\CR)$ & 10363 & 48537 & 0.20 & 31.64 & \textbf{31.84} & 385.19 & timeout & \ding{51} & \ding{51} \\   \midrule
        CRAFT-R4-b3-D & $\zeta(3,1,\mathcal{T},\CR)$ & 13516 & 63332 & 0.61 & 61.46 & \textbf{62.07} & 1188.03 & timeout & \ding{51}  & \ding{51}\\
        CRAFT-R4-b3-D & $\zeta(4,1,\mathcal{T},\CR)$ & 14039 & 67789 & 0.63 & 0.14 & \textbf{0.77} & 8.25 & timeout & \myding{55} & \myding{55}\\
        CRAFT-R4-b3-D & $\zeta(3,2,\mathcal{T},\CR)$ & 13518 & 63363 & 0.61 & 155.14 & \textbf{155.75} & 978.78 & timeout & \ding{51} & \ding{51} \\
        CRAFT-R4-b3-D & $\zeta(3,3,\mathcal{T},\CR)$ & 13518 & 63363 & 0.61 & 81.47 & \textbf{82.08} & 900.28 & timeout & \ding{51}  & \ding{51}\\
        CRAFT-R4-b3-D & $\zeta(3,4,\mathcal{T},\CR)$ & 13516 & 63332 & 0.62 & 70.53 & \textbf{71.15} & 916.13 & timeout & \ding{51}  & \ding{51}\\
        CRAFT-R4-b3-D & $\zeta(3,5,\mathcal{T},\CR)$ & 13515 & 63325 & 0.61 & 54.55 & \textbf{55.16} & 922.49 & timeout & \ding{51}  & \ding{51}\\   \midrule
        LED64-R1-b1-D & $\zeta(1,1,\mathcal{T},\CR)$ & 2673 & 13295 & 0.05 & 1.07 & 1.12 & 11.63 & \textbf{0.23} & \ding{51}  & \ding{51}\\
        LED64-R1-b1-D & $\zeta(2,1,\mathcal{T},\CR)$ & 3038 & 12073 & 0.05 & 0.02 & \textbf{0.07} & 0.85 & 8.39 & \myding{55} & \myding{55}  \\
        LED64-R1-b2-D & $\zeta(2,1,\mathcal{T},\CR)$ & 3726 & 16057 & 0.08 & 2.01 & \textbf{2.09} & 33.40 & 10.18 & \ding{51}  & \ding{51}\\
        LED64-R1-b2-D & $\zeta(3,1,\mathcal{T},\CR)$ & 3853 & 16852 & 0.09 & 0.02 & \textbf{0.11} & 0.83 & 2292.60 & \myding{55}  & \myding{55} \\   \midrule
        LED64-R2-b1-D & $\zeta(1,1,\mathcal{T},\CR)$ & 5215 & 27405 & 0.12 & 9.64 & \textbf{9.76} & 139.78 & timeout & \ding{51}& \ding{51}  \\
        LED64-R2-b1-D & $\zeta(2,1,\mathcal{T},\CR)$ & 5783 & 26245 & 0.12 & 0.05 & \textbf{0.17} & 1.68 & timeout & \myding{55}  & \myding{55} \\
        LED64-R2-b2-D & $\zeta(2,1,\mathcal{T},\CR)$ & 7300 & 34132 & 0.18 & 13.67 & \textbf{13.85} & 336.45 & timeout & \ding{51} & \ding{51} \\
        LED64-R2-b2-D & $\zeta(3,1,\mathcal{T},\CR)$ & 7554 & 36930 & 0.18 & 0.06 & \textbf{0.24} & 2.78 & timeout & \myding{55}& \myding{55}  \\   \midrule
        LED64-R3-b1-D & $\zeta(1,1,\mathcal{T},\CR)$ & 7674 & 40655 & 0.14 & 36.57 & \textbf{36.71} & 588.07 & timeout & \ding{51} & \ding{51} \\
        LED64-R3-b1-D & $\zeta(2,1,\mathcal{T},\CR)$ & 8528 & 38988 & 0.14 & 0.11 & \textbf{0.25} & 2.90 & timeout & \myding{55} & \myding{55} \\
        LED64-R3-b2-D & $\zeta(2,1,\mathcal{T},\CR)$ & 10924 & 49866 & 0.24 & 47.75 & \textbf{47.99} & 860.20 & timeout & \ding{51} & \ding{51} \\
        LED64-R3-b2-D & $\zeta(3,1,\mathcal{T},\CR)$ & 11305 & 53487 & 0.25 & 0.12 & \textbf{0.37} & 3.72 & timeout & \myding{55} & \myding{55} \\   \bottomrule
    \end{tabular}}\vspace{-3mm}
\end{table}

\subsection{RQ1: Efficiency and Effectiveness of \tool for Fault-resistance Verification}
To answer RQ1, we compare  \tool with (1) the state-of-the-art verifier FIVER~\cite{RichterBrockmann2021FIVERR} and (2) an SMT-based approach which directly
checks the constraints generated by our encoding method without translating to Boolean formulas.
We use the SMT solver bitwuzla~\cite{NPre20}, which is the winner of QF\_Bitvec (Single Query Track) Division at SMT-COMP 2021 and 2022.

The results are reported in Table~\ref{tab:fault-resistancemf}, where both the SAT solver Glucose and the BDD-based verifier FIVER are run with 8 threads
while the SMT solver bitwuzla is run with a single thread, because there is no promising parallel SMT solvers for (quantifier-free) bit-vector theory.
Columns (2CNF) and (Solving) give the execution time of building and solving Boolean formulas  in seconds, respectively.
Columns (Total) and (Time) give the total execution time in seconds.
Mark \ding{51} (resp. \myding{55}) indicates that the protected circuit $\SeqC'$ is fault-resistant (resp. not fault-resistant).

Overall, both \tool (i.e., the SAT-based approach) and the SMT-based approach solved all the verification tasks, while FIVER runs out of time on 23 verification tasks within the time limit (24 hours per task).
The SAT/SMT-based approach become more and more efficient than FIVER
with the increase of round numbers (i.e., R$i$) and maximal number of protected faulty bits (i.e., b$j$).
\tool is significantly more efficient than the SMT-based approach on relatively larger benchmarks (e.g., AES-R1-b1, AES-R1-b2, CRAFT-R1-b2-C, CRAFT-R2-b2-C, CRAFT-R3-b3-D, CRAFT-R4-b3-D, LED64-R2-b1-D, LED64-R2-b2-D, LED64-R3-b1-D and LED64-R3-b2-D) while they are comparable on smaller benchmarks.

Interestingly, we found that
(i)  implementations with correction-based countermeasures are more difficult to prove than that with detection-based countermeasures (e.g., CRAFT-R$i$-b$j$-C vs. CRAFT-R$i$-b$j$-D, for $i=1,2$ and $j=1,2$), because implementing correction-based countermeasures require more gates;
(ii) \tool is more efficient at disproving fault-resistance than proving fault-resistance, because UNSAT instances are often more difficult to prove than SAT instances in CDCL SAT solvers.
(iii) \tool often scales very well with increase of the round numbers (i.e., R$i$ for $i=1,2,3,4$),
maximal number of protected faulty bits (i.e., b$j$ for $j=1,2,3$), maximum number of fault events per clock cycle (i.e., $\Nn_e$)
and the maximum number of clock cycles in which fault events can occur  (i.e., $\Nn_c$),
but FIVER has very limited scalability.

\begin{figure}[t]
    \centering
    \subfigure[AES-R1-b1-D and AES-R1-b2-D]{
    \label{fig:AESthreadsmix}
    \includegraphics[scale=0.285]{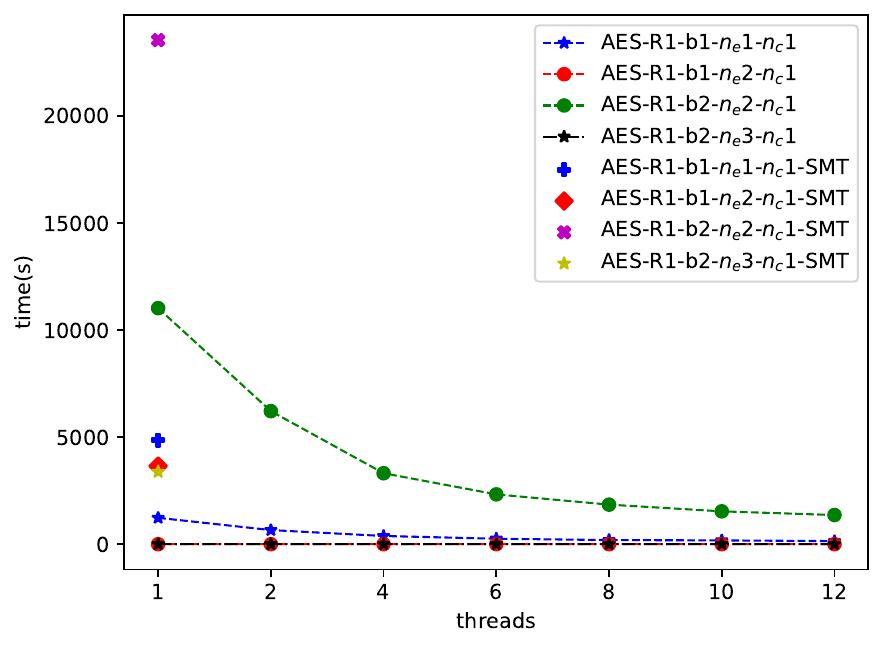}}
    \subfigure[CRAFT-R4-b3-D]{
    \label{fig:CRAFTthreadsmix}
    \includegraphics[scale=0.285]{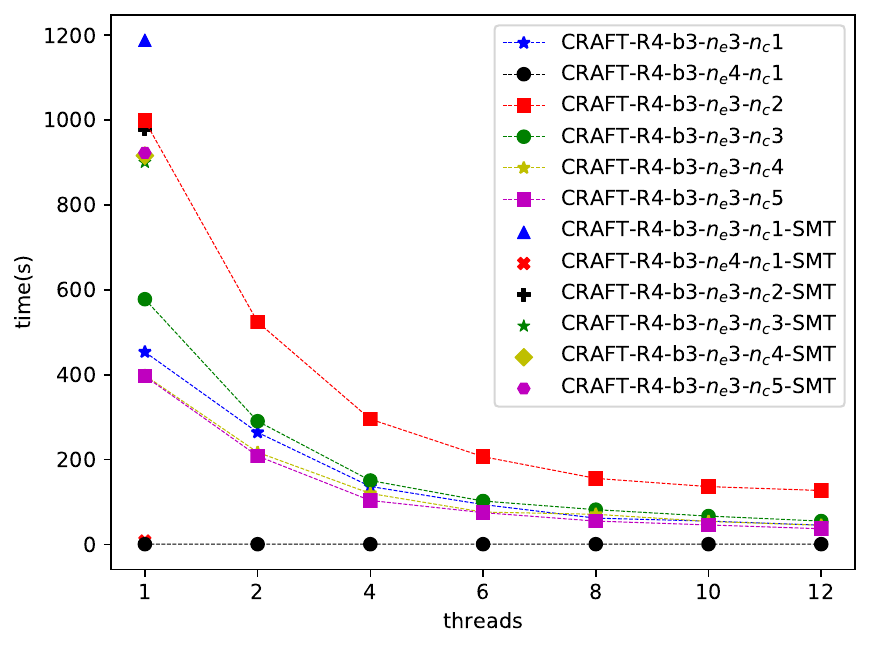}}\vspace{-1mm}
    \caption{Comparison of verification times with different number of threads}\vspace{-2mm}
\end{figure}

To understand the effect of the number of threads, we evaluate \tool on the benchmarks AES-R1-b1-D, AES-R1-b2-D, and CRAFT-R4-b3-D
by varying the number of threads from 1 to 12. The results are depicted in Fig.~\ref{fig:AESthreadsmix}
and Fig.~\ref{fig:CRAFTthreadsmix}, respectively, where $\Nn_ei$ and $\Nn_cj$ denote the fault-resistance mode $\zeta(i,j,\mathcal{T},\CR)$.
Detailed results are reported in the appendix. 
We observe that \tool almost always outperforms the SMT-based approach.
On the fault-resistant benchmarks (i.e., curves with b$j$-$\Nn_e k$ such that $j\geq k$), \tool becomes more and more efficient
while the improvement becomes less and less, with the increase of the number of threads.
On the non-fault-resistant benchmarks (i.e., curves with b$j$-$\Nn_e k$ such that $j<k$), multi-threading does not improve performance
instead slightly worsens performance, because they are easy to be disproved and thread scheduling causes overhead.

\begin{tcolorbox}[size=title]
	{\textbf{Answer to RQ1:}}
\tool is able to efficiently and effectively verify the fault-resistance of realistic cryptographic
circuits with detection- and correction-based countermeasures.
It performs significantly better than the state-of-the-art, as well as an SMT-based appraoch.
\end{tcolorbox}

\subsection{RQ2: Effectiveness of the Vulnerable Gate Reduction}
To answer RQ2, we evaluate \tool with/without the vulnerable gate reduction using 8 threads for SAT solving,
considering all the fault types $\mathcal{T} = \{ \tau_{s},\tau_{r},\tau_{bf}\}$.

The results are reported in Table~\ref{tab:faultgatereductionmixed},
where columns (\#Gate) give the number of vulnerable gates that should be considered when verifying
fault-resistance. We can observe that our vulnerable gate reduction is able to significantly reduce the number of vulnerable gates that should be considered when verifying
fault-resistance, achieving more than 72\% reduction rate on average, consequently, significantly reduce the size of the resulting Boolean formulas.
Interestingly, reducing the size of the resulting Boolean formulas does not necessarily improve the overall verification efficiency.
Indeed, the vulnerable gate reduction is very effective in proving fault-resistant benchmarks no matter the adopted countermeasure, fault-resistance model,
fault-resistance and size of the benchmarks, but sightly worsens the performance for disproving non-fault-resistant benchmarks as they are easy to disprove
and the vulnerable gate reduction itself has some overhead.

\begin{tcolorbox}[size=title]
	{\textbf{Answer to RQ2:}}
Our vulnerable gate reduction achieves on average more than 72\% reduction rate of vulnerable gates and is very effective in proving fault-resistant benchmarks.
\end{tcolorbox}

The results w.r.t. the fault-resistance models limited to the fault type $\tau_{bf}$ are given in the appendix 
from which a similar conclusion can be drawn.

\begin{table}[t]
\centering\setlength\tabcolsep{1pt}
  \caption{Results of fault-resistance verification using \tool with/without vulnerable gate reduction.}\vspace{-1mm}
  \label{tab:faultgatereductionmixed}
\scalebox{0.62}{
    \begin{tabular}{l|c| rrrr| rrrr|c|c}    \toprule
  \multirow{2}{*}{\bf Name} & \textbf{Fault-resistance} & \multicolumn{4}{c|}{\bf Without vulnerable gate reduction}  & \multicolumn{4}{c|}{\bf With vulnerable gate reduction}  &  \multirow{2}{*}{\bf Result} & {\bf Desired}  \\ \cline{3-10}
                             & \textbf{model} &  \textbf{\#Var} & \textbf{\#Clause} & \textbf{\#Gate}  & \textbf{Time} & \textbf{\#Var} & \textbf{\#Clause} & \textbf{\#Gate} & \textbf{Time} &  &{\bf Result} \\  \midrule
        AES-R1-b1-D   & $\zeta (1,1,\mathcal{T},\CR)$ & 149993 & 749449 & 24432 & 2334.24 & 57340 & 310569 & \textbf{7920} & \textbf{195.24} & \ding{51} & \ding{51} \\
        AES-R1-b1-D   & $\zeta (2,1,\mathcal{T},\CR)$ & 172910 & 589178 & 24432 & 2.74 & 69088 & 246108 & \textbf{7920} & \textbf{2.57} & \myding{55}& \myding{55}  \\
        AES-R1-b2-D   & $\zeta (2,1,\mathcal{T},\CR)$ & 247606 & 698415 & 33104 & timeout & 86333 & 329543 & \textbf{10640} & \textbf{1850.07} & \ding{51}& \ding{51}  \\
        AES-R1-b2-D   & $\zeta (3,1,\mathcal{T},\CR)$ & 263989 & 748906 & 33104 & \textbf{5.10} & 90428 & 354018 & \textbf{10640} & 5.28 & \myding{55} & \myding{55} \\    \midrule
        CRAFT-R1-b1-C & $\zeta (1,1,\mathcal{T},\CR)$ & 19233 & 92262 & 2948 & 4.24 & 6767 & 33938 & \textbf{760} & \textbf{0.79} & \ding{51} & \ding{51} \\
        CRAFT-R1-b1-C & $\zeta (2,1,\mathcal{T},\CR)$ & 21882 & 82851 & 2948 & \textbf{0.15} & 7608 & 32039 & \textbf{760} & 0.20 & \myding{55} & \myding{55} \\
        CRAFT-R1-b2-C & $\zeta (2,1,\mathcal{T},\CR)$ & 148552 & 558649 & 19636 & 2519.00 & 52901 & 234385 & \textbf{5672} & \textbf{132.79} & \ding{51}& \ding{51}  \\
        CRAFT-R1-b2-C & $\zeta (3,1,\mathcal{T},\CR)$ & 157002 & 642847 & 19636 & \textbf{2.84} & 55099 & 258959 & \textbf{5672} & 3.39 & \myding{55} & \myding{55} \\
        CRAFT-R1-b1-D & $\zeta (1,1,\mathcal{T},\CR)$ & 5185	& 23589 & 766 & 0.30 & 2255 & 10735 & \textbf{274} & \textbf{0.13} & \ding{51} & \ding{51} \\
        CRAFT-R1-b1-D & $\zeta (2,1,\mathcal{T},\CR)$ & 6021	& 21519 & 766 & \textbf{0.05} & 2567 & 10316 & \textbf{274} & 0.06 & \myding{55}& \myding{55}  \\
        CRAFT-R1-b2-D & $\zeta (2,1,\mathcal{T},\CR)$ & 8402	& 31227 & 1139 & 1.64 & 3356 & 13780 & \textbf{376} & \textbf{0.24} & \ding{51} & \ding{51} \\
        CRAFT-R1-b2-D & $\zeta (3,1,\mathcal{T},\CR)$ & 8900	& 35991 & 1139 & 0.07 & 3538 & 15058 & \textbf{376} & \textbf{0.06} & \myding{55} & \myding{55} \\
        CRAFT-R1-b3-D & $\zeta (3,1,\mathcal{T},\CR)$ & 10236 & 41474 & 1296 & 7.62 & 3974 & 19042 & \textbf{448} & \textbf{0.46} & \ding{51} & \ding{51} \\
        CRAFT-R1-b3-D & $\zeta (4,1,\mathcal{T},\CR)$ & 10714 & 45532 & 1296 & 0.07 & 4132 & 20428 & \textbf{448} &  0.07 & \myding{55} & \myding{55} \\   \midrule
        CRAFT-R2-b1-C & $\zeta (1,1,\mathcal{T},\CR)$ & 36565 & 176024 & 5656 & 60.97 & 12644 & 63451 & \textbf{1440} & \textbf{11.63} & \ding{51} & \ding{51} \\
        CRAFT-R2-b1-C & $\zeta (2,1,\mathcal{T},\CR)$ & 42083 & 157194 & 5656 & \textbf{0.34} & 14215 & 59873 & \textbf{1440} & 0.45 & \myding{55}& \myding{55}  \\
        CRAFT-R2-b2-C & $\zeta (2,1,\mathcal{T},\CR)$ & 291588 & 1099486	& 38872 & timeout & 104139 & 452464 & \textbf{11200} & \textbf{2060.81} & \ding{51} & \ding{51} \\
        CRAFT-R2-b2-C & $\zeta (3,1,\mathcal{T},\CR)$ & 308229 & 1269539	& 38872 & \textbf{7.80} & 108384 & 498341 & \textbf{11200} & 10.41 & \myding{55}& \myding{55}  \\
        CRAFT-R2-b1-D & $\zeta (1,1,\mathcal{T},\CR)$ & 9960	& 45624 & 1480 & 4.98 & 4195 & 20279 & \textbf{508} & \textbf{0.94} & \ding{51}& \ding{51}  \\
        CRAFT-R2-b1-D & $\zeta (2,1,\mathcal{T},\CR)$ & 11849 & 41318 & 1480 & \textbf{0.06} & 4795 & 19338 & \textbf{508} &  0.10 & \myding{55} & \myding{55} \\
        CRAFT-R2-b1-D & $\zeta (1,2,\mathcal{T},\CR)$ & 9960	& 45624 & 1480 & 7.22 & 4195 & 20279 & \textbf{508} & \textbf{1.82} & \ding{51} & \ding{51} \\
        CRAFT-R2-b1-D & $\zeta (1,3,\mathcal{T},\CR)$ & 9959	& 45622 & 1480 & 6.35 & 4194 & 20277 & \textbf{508} & \textbf{1.41} & \ding{51}& \ding{51}  \\
        CRAFT-R2-b2-D & $\zeta (2,1,\mathcal{T},\CR)$ & 16137 & 59108 & 2180 & 60.37 & 5959 & 25370 & \textbf{672} & \textbf{3.34} & \ding{51} & \ding{51} \\
        CRAFT-R2-b2-D & $\zeta (3,1,\mathcal{T},\CR)$ & 17146 & 67617 & 2180 & \textbf{0.09} & 6268 & 27935 & \textbf{672} & 0.17 & \myding{55} & \myding{55} \\
        CRAFT-R2-b2-D & $\zeta (2,2,\mathcal{T},\CR)$ & 16137 & 59108 & 2180 & 91.31 & 5959 & 25370 & \textbf{672} & \textbf{3.58} & \ding{51} & \ding{51} \\   \midrule
        CRAFT-R3-b3-D & $\zeta (3,1,\mathcal{T},\CR)$ & 29109 & 119401 & 3776 & 7049.02	& 10364 & 48541 & \textbf{1104} & \textbf{37.86} & \ding{51}& \ding{51}  \\
        CRAFT-R3-b3-D & $\zeta (4,1,\mathcal{T},\CR)$ & 30481 & 131197 & 3776 & \textbf{0.25} & 10760 & 51937 & \textbf{1104} & 0.31 & \myding{55} & \myding{55} \\
        CRAFT-R3-b3-D & $\zeta (3,2,\mathcal{T},\CR)$ & 29111 & 119420 & 3776 & 3547.68 & 10366 & 48560 & \textbf{1104} & \textbf{57.74} & \ding{51}& \ding{51}  \\
        CRAFT-R3-b3-D & $\zeta (3,3,\mathcal{T},\CR)$ & 29109 & 119401 & 3776 & 2361.28 & 10364 & 48541 & \textbf{1104} & \textbf{37.04} & \ding{51}& \ding{51}  \\
        CRAFT-R3-b3-D & $\zeta (3,4,\mathcal{T},\CR)$ & 29108 & 119397 & 3776 & 1513.75	& 10363 & 48537 & \textbf{1104} & \textbf{31.84} & \ding{51} & \ding{51} \\   \midrule
        CRAFT-R4-b3-D & $\zeta (3,1,\mathcal{T},\CR)$ & 38299 & 158236 & 4992 & 6074.49	& 13516 & 63332 & \textbf{1440} & \textbf{62.07} & \ding{51} & \ding{51} \\
        CRAFT-R4-b3-D & $\zeta (4,1,\mathcal{T},\CR)$ & 40118 & 173925 & 4992 & \textbf{0.76} & 14039 & 67789 & \textbf{1440} & 0.77 & \myding{55} & \myding{55} \\
        CRAFT-R4-b3-D & $\zeta (3,2,\mathcal{T},\CR)$ & 38301 & 158267 & 4992 & 15014.46 & 13518 & 63363 & \textbf{1440} & \textbf{155.75} & \ding{51} & \ding{51} \\
        CRAFT-R4-b3-D & $\zeta (3,3,\mathcal{T},\CR)$ & 38301 & 158267 & 4992 & 5448.20	& 13518 & 63363 & \textbf{1440} & \textbf{82.08} & \ding{51} & \ding{51} \\
        CRAFT-R4-b3-D & $\zeta (3,4,\mathcal{T},\CR)$ & 38299 & 158236 & 4992 & 4292.37	& 13516 & 63332 & \textbf{1440} & \textbf{71.15} & \ding{51} & \ding{51} \\
        CRAFT-R4-b3-D & $\zeta (3,5,\mathcal{T},\CR)$ & 38298 & 158229 & 4992 & 4009.42	& 13515 & 63325 & \textbf{1440} & \textbf{55.16} & \ding{51} & \ding{51} \\   \midrule
        LED64-R1-b1-D & $\zeta (1,1,\mathcal{T},\CR)$ & 8306	& 39057 & 1312 & 4.81 & 2673 & 13295 & \textbf{240} & \textbf{1.12} & \ding{51} & \ding{51} \\
        LED64-R1-b1-D & $\zeta (2,1,\mathcal{T},\CR)$ & 9749	& 31481 & 1312 & 0.12 & 3038 & 12073 & \textbf{240} & \textbf{0.07} & \myding{55}& \myding{55}  \\
        LED64-R1-b2-D & $\zeta (2,1,\mathcal{T},\CR)$ & 14433 & 41129 & 1888 & 24.79 & 3726 & 16057 & \textbf{336} & \textbf{2.09} & \ding{51}& \ding{51}  \\
        LED64-R1-b2-D & $\zeta (3,1,\mathcal{T},\CR)$ & 15296 & 43876 & 1888 & 0.18 & 3853 & 16852 & \textbf{336} & \textbf{0.11} & \myding{55} & \myding{55} \\   \midrule
        LED64-R2-b1-D & $\zeta (1,1,\mathcal{T},\CR)$ & 16464 & 79748 & 2496 & 65.85 & 5215 & 27405 & \textbf{480} & \textbf{9.76} & \ding{51} & \ding{51} \\
        LED64-R2-b1-D & $\zeta (2,1,\mathcal{T},\CR)$ & 18527 & 72422 & 2496 & 0.25 & 5783 & 26245 & \textbf{480} & \textbf{0.17} & \myding{55} & \myding{55} \\
        LED64-R2-b2-D & $\zeta (2,1,\mathcal{T},\CR)$ & 27208 & 95236 & 3536 & 170.95 & 7300 & 34132 & \textbf{672} & \textbf{13.85} & \ding{51} & \ding{51} \\
        LED64-R2-b2-D & $\zeta (3,1,\mathcal{T},\CR)$ & 28694 & 106194 & 3536 & 0.38 & 7554 & 36930 & \textbf{672} & \textbf{0.24} & \myding{55} & \myding{55} \\   \midrule
        LED64-R3-b1-D & $\zeta (1,1,\mathcal{T},\CR)$ & 24168 & 117662 & 3680 & 233.72 & 7674 & 40655 & \textbf{720} & \textbf{36.71} & \ding{51} & \ding{51} \\
        LED64-R3-b1-D & $\zeta (2,1,\mathcal{T},\CR)$ & 27669 & 105378 & 3680 & \textbf{0.21} & 8528 & 38988 & \textbf{720} & 0.25 & \myding{55}& \myding{55}  \\
        LED64-R3-b2-D & $\zeta (2,1,\mathcal{T},\CR)$ & 39256 & 140834 & 5184 & 343.16 & 10924 & 49866 & \textbf{1008} & \textbf{47.99} & \ding{51} & \ding{51} \\
        LED64-R3-b2-D & $\zeta (3,1,\mathcal{T},\CR)$ & 41301 & 158823 & 5184 & \textbf{0.31} & 11305 & 53487 & \textbf{1008} & 0.37 & \myding{55}& \myding{55}  \\   \bottomrule
    \end{tabular}}
\end{table}

\section{Related work}
\label{chap:related work}
This work focuses on formal verification
of countermeasures against fault injection attacks.
In this section, we discuss related work on functional equivalence checking, safety and fault-resistance verification of hardware designs.

Functional equivalence checking of hardware designs can be roughly divided into
combinational and sequential equivalence checking~\cite{kuehlmann2002combinational},
where the former requires that the given gates match in the circuits under the same inputs,
while the latter only requires that the outputs match in the circuits under the same inputs.
Various combinational and sequential equivalence checking techniques have been proposed such as SAT/SMT-based ones (e.g.,~\cite{GoldbergPB01,KhasidashviliH03,KaissSHK07,AzarbadA17,MishchenkoCBE06})
and BDD-based ones (e.g.,~\cite{Pixley92,Eijk98,Bryant86,KuehlmannK97}). 
Safety verification of hardware designs is usually done by model-checking, where safety properties are expressed as assertions using temporal logic.
Both SAT/SMT-based (e.g., \cite{BiereCCZ99,MukherjeeSKM16,JainKSC08,BiereCCFZ99,BruttomessoCFGHNPS07,BraytonM10,Bradley11,LeeS14}) and BDD-based (e.g.,~\cite{BurchCMDH90,BurchCLMD94,chaki2018bdd}) methods have been widely studied. 
However, all the approaches and tools for checking functional equivalence and safety properties cannot be \emph{directly} applied
to check fault-resistance, though our SAT encoding method is inspired by the existing SAT-based ones for checking sequential equivalence.
Indeed, the fault-resistance problem is significantly different from 
the functional equivalence problem and cannot be easily expressed as a safety property.
Mover, our vulnerable gate reduction which is every effective in improving verification efficiency
cannot be leveraged using existing tools.

Due to the severity of fault injection attacks, many simulation- and SAT-based approaches have been proposed to find effective fault vectors
or check the effectiveness of a user-specified fault vector (e.g.,~\cite{SimevskiKK13,BurchardGEH00KP17,saha2018expfault,arribas2020cryptographic,khanna2017xfc,srivastava2020solomon,wang2021sofi,nasahl2022synfi,BDFGZ14,FujitaM14,WangMGF18,WangGF19,WangGF20}).
Though promising, it is infeasible if not impossible to verify a given hardware design considering all possible fault vectors that could occur under all valid input combinations.
To fill this gap,
a BDD-based approach, named FIVER, was proposed~\cite{RichterBrockmann2021FIVERR}, which exclusively focuses on fault-resistance verification.
In general, for each possible fault vector, FIVER builds a BDD model to represent the concrete faulty circuit w.r.t. the fault vector,
and analyzes fault-resistance w.r.t. the fault vector by comparing the BDD model with the BDD model of the original circuit.
Though several optimizations were proposed, it suffers from the combinatorial exploration problem with the increase of fault types, vulnerable gates and clock cycles,
thus is limited in efficiency and scalability.
Thanks to our novel fault event encoding method and effective vulnerable gate reduction, our approach does not need to explicitly enumerate all the possible fault vectors
and the verification process can fully utilize the conflict-driven clause learning feature of modern SAT solvers.
Thus, \tool scales very well and is significantly more efficient than FIVER on relatively larger benchmarks.
 %

Countermeasure synthesis techniques  have also been proposed to repair flaws (e.g.,~\cite{EldibWW16,wang2021sofi,RoyRHB20}).
However, they do not provide security guarantees (e.g.,~\cite{wang2021sofi,RoyRHB20}) or
are limited to one specific type of fault injection attacks (e.g., clock glitches in ~\cite{EldibWW16}) and thus are still vulnerable to other fault injection attacks.

\section{Conclusion}
\label{chap:conclusion}
We formalized the fault-resistance verification problem of cryptographic circuits and proved that it is NP-complete for the first time. 
We proposed novel fault encoding and SAT encoding methods to reduce the fault-resistance verification problem into the SAT problem so that state-of-the-art SAT solvers can be harnessed.
We also presented a novel vulnerable gate reduction technique to effectively reduce the number of vulnerable gates, which can significantly improve the verification efficiency.
We implemented our approach in an open-source tool and extensively evaluate it on a
set of realistic cryptographic circuits. 
Experimental results show that our approach  significantly outperforms the state-of-the-art
for fault-resistance verification.
Our tool enables hardware designers to assess and verify the
countermeasures in a systematic and automatic way.

For future research, it would be interesting to develop automated flaw repair techniques by leveraging the
verification results produced by our approach.



%


\bibliographystyle{IEEEtranS}
\bibliography{refs}
\appendix


\section{Missing Proofs}
\label{sec:proofs}

\subsection{Proof of Theorem~\ref{thm:fault-resistance-NP}}

{\sc Theorem~\ref{thm:fault-resistance-NP}}.
\textit{The problem of determining whether a circuit $\SeqC'$ is not fault-resistant is
NP-complete. }

\begin{proof}
To show that the problem is in NP, we can first non-deterministically guess a sequence of primary inputs $(\vec{x}_{1},\cdots,\vec{x}_{k})$ and a fault vector $\VV(\SeqC',\BB,T)\in \sem{\zeta(\Nn_e,\Nn_c,T,\ell)}$,
then construct the faulty circuit $\SeqC'[\VV(\SeqC',\BB,T)]$ in polynomial time,
finally compute and check if the sequences of primary outputs $\sem{\SeqC'}(\vec{x}_1,\cdots, \vec{x}_k)$ and $\sem{\SeqC'[\VV(\SeqC',\BB,T)]}(\vec{x}_1,\cdots, \vec{x}_k)$
differ at some clock cycle before the error flag output $o_{\tt flag}$ differs in polynomial time.
If not, then  $\langle\SeqC',\BB\rangle\not\models \zeta(\Nn_e,\Nn_c,T,\ell)$.

\begin{figure}[t]
  \centering
  \includegraphics[width=0.48\textwidth]{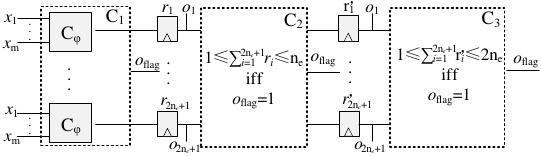}\\
  \caption{The circuit $\SeqC'$ for NP-hardness.}\label{fig:NPhard}
\end{figure}

The NP-hardness is proved by reducing from the SAT problem.
Let $C_\varphi$ be a combinational circuit representing a Boolean formula $\varphi$, where the inputs of $C_\varphi$ are the Boolean variables
of $\varphi$, and the output indicates the result of $\varphi$.
We create a circuit $\SeqC'=(\II,\OO,\RR, \vec{s}_0, \CC)$ as shown in Fig.~\ref{fig:NPhard}, where
\begin{itemize}
  \item $\II=\{x_1,\cdots,x_m\}$ is the set of inputs of the circuit $C_\varphi$;
  \item $\OO$ is the set $\{o_i,o_{\tt flag}\mid 1\leq i\leq 2\Nn_e+1\}$;
  \item $\RR=R_1\cup R_2$, where $R_1=\{r_i\mid 1\leq i\leq 2\Nn_e+1\}$ and $R_2=\{r_i'\mid 1\leq i\leq 2\Nn_e+1\}$;
  \item $\vec{s}_0$ is a vector consisting of $\Gzero$;
  \item $\CC=\{C_1,C_2,C_3\}$, where
  \begin{itemize}
  \item $C_1$ comprises $2\Nn_e+1$ copies of the circuit $C_\varphi$: all the copies share the same inputs $\II$, the output
  of the $i$-th copy is connected to $r_i$, and the output $o_{\tt flag}$ is always $\Gzero$;
  \item $C_2$ outputs signals of the memory gates $R_1$ and store them into the memory gates $R_2$ again,
  checks if $1\leq \sum_{i=1}^{2\Nn_e+1} r_i\leq \Nn_e$, and the output $o_{\tt flag}$ is $\Gone$ iff $1\leq \sum_{i=1}^{2\Nn_e+1} r_i\leq \Nn_e$;
  \item $C_3$ checks whether $1\leq \sum_{i=1}^{2\Nn_e+1} r_i'\leq 2\Nn_e$, and the output $o_{\tt flag}$ is $\Gone$ iff $1\leq \sum_{i=1}^{2\Nn_e+1} r_i'\leq 2\Nn_e$.
  \end{itemize}
\end{itemize}

\noindent
{\bf Claim}. \textit{The circuit $\SeqC'$ is not fault-resistant w.r.t. the blacklist $\BB=\emptyset$ and the fault-resistance model $\zeta(\Nn_e,1,\{\tau_{bf}\},\Rr)$ iff
the Boolean formula $\varphi$ is satisfiable.}

$(\Leftarrow)$ Suppose $\varphi$ is satisfiable. Let $\vec{x}$ be the satisfying assignment of $\varphi$.
Obviously, under the primary inputs $\vec{x}$, the output $o_{\tt flag}$ is $\Gzero$ and the outputs $\{o_i\mid 1\leq i\leq 2\Nn_e+1\}$ are $\Gone$ in all the clock cycles.
Consider the fault event $\EE(2,r_1,\tau_{bp})$.
Along the sequence of primary outputs $\sem{\SeqC'[\EE(2,r_1,\tau_{bp})]}(\vec{x})$,
the output $o_{\tt flag}$ is $\Gzero$ at the first two clock cycles and becomes $\Gone$ at the 3-rd clock cycle.
However, the output $o_1$ differs in $\sem{\SeqC'}(\vec{x})$ and $\sem{\SeqC'[\EE(2,r_1,\tau_{bp})]}(\vec{x})$
at the 2-nd clock cycle due to the bit-flip fault injection on the memory gate
$r_1$. Thus,  $\langle\SeqC',\emptyset\rangle\not\models \zeta(\Nn_e,1,\{\tau_{bf}\},\Rr)$.


$(\Rightarrow)$ Suppose $\varphi$ is unsatisfiable. 
Obviously, under any primary inputs $\vec{x}$, all the primary outputs $\{o_{\tt flag},o_i\mid 1\leq i\leq 2\Nn_e+1\}$ are $\Gzero$ in all the clock cycles.
For any fault vector $\VV(\SeqC',\BB,T)\in\zeta(\Nn_e,1,\{\tau_{bf}\},\Rr)$,
at most $\Nn_e$ memory gates can be bit-flipped in one single clock cycle. If
some memory gates in $R_1$ are bit-flipped at the 2-nd clock cycle, then the output $o_{\tt flag}$ is $\Gone$ at the 2-nd clock cycle.
If no memory gates of $R_1$ are bit-flipped at the 2-nd clock cycle and some memory gates in $R_2$ are bit-flipped at the 3-rd clock cycle,
the primary outputs $\{o_i\mid 1\leq i\leq 2\Nn_e+1\}$ are $\Gzero$ at the 2-nd clock cycle,
$o_{\tt flag}$ is $\Gone$ at the 3-nd clock cycle although some primary outputs of $\{o_i\mid 1\leq i\leq 2\Nn_e+1\}$
become $\Gone$ at the 3-nd clock cycle. Thus,  $\langle\SeqC',\emptyset\rangle\models \zeta(\Nn_e,1,\{\tau_{bf}\},\Rr)$
%
\end{proof}

\subsection{Proof of Theorem~\ref{thm:faultgatereduce}}

{\sc Theorem~\ref{thm:faultgatereduce}}.
\textit{Consider a fault-resistance model $\zeta(\Nn_e,\Nn_c,T,\ell)$ such that $\tau_{bf}\in T$ or $\{\tau_s,\tau_r\}\subseteq T$, and $\ell\in\{\Cc,\CR\}$.
Let $\VV_1(\SeqC',\BB,T)=\VV(\SeqC',\BB,T)\cup \{\EE(\alpha,\beta,\tau)\}\in \sem{\zeta(\Nn_e,\Nn_c,T,\ell)}$ be an effective fault vector on the circuit $\SeqC'$.
If the output of the gate $\beta$ is only connected to one logic gate $\beta'\not\in\BB$, 
then 
there exists a fault vector $\VV'(\SeqC',\BB,T)\subseteq \VV(\SeqC',\BB,T)\cup \{\EE(\alpha,\beta',\tau')\}$ for some $\tau'\in T$ such
that $\VV'(\SeqC',\BB,T)$ is also effective on the circuit $\SeqC'$.}

\textit{Moreover, if $\VV(\SeqC',\BB,T)=\emptyset$, then $\{\EE(\alpha,\beta',\tau')\}$ for some $\tau'\in T$ is effective on the circuit $\SeqC'$.}

\begin{proof}%

Since $\VV_1(\SeqC',\BB,T)$ is an effective fault vector on the circuit $\SeqC'$,
there exists a sequence of primary inputs $(\vec{x}_{1},\cdots,\vec{x}_{k})$ such that
 $\sem{\SeqC'}(\vec{x}_1,\cdots, \vec{x}_k)$
and $\sem{\SeqC'[\VV_1(\SeqC',\BB,T)]}(\vec{x}_1,\cdots, \vec{x}_k)$ differ at
some clock cycle before the error flag output $o_{\tt flag}$ differs.
We proceed by distinguishing whether the output signal of the gate $\beta'$ differs in the circuits
$\SeqC'$ and $\SeqC'[\VV_1(\SeqC',\BB,T)]$ under the same sequence of primary inputs $(\vec{x}_{1},\cdots,\vec{x}_{k})$.

\begin{itemize}
  \item If the output signal of the gate $\beta'$ is the same in the circuits
$\SeqC'$ and $\SeqC'[\VV_1(\SeqC',\BB,T)]$ under the same sequence of primary inputs $(\vec{x}_{1},\cdots,\vec{x}_{k})$,
then the effect of the fault event $\EE(\alpha,\beta,\tau)$ is stopped at the gate $\beta'$, as
the output of the gate $\beta$ is only connected to the gate $\beta'$. Thus,
the sequences of primary outputs $\sem{\SeqC'[\VV(\SeqC',\BB,T)]}(\vec{x}_1,\cdots, \vec{x}_k)$
and $\sem{\SeqC'[\VV_1(\SeqC',\BB,T)]}(\vec{x}_1,\cdots, \vec{x}_k)$ are the same.
It implies that  $\sem{\SeqC'}(\vec{x}_1,\cdots, \vec{x}_k)$
and $\sem{\SeqC'[\VV(\SeqC',\BB,T)]}(\vec{x}_1,\cdots, \vec{x}_k)$ differ at
some clock cycle before the error flag output $o_{\tt flag}$ differs.
The result immediately follows.

\item If the output signal of the gate $\beta'$ differs in the circuits
$\SeqC'$ and $\SeqC'[\VV_1(\SeqC',\BB,T)]$ under the same sequence of primary inputs $(\vec{x}_{1},\cdots,\vec{x}_{k})$,
then the fault propagation from the fault event $\EE(\alpha,\beta,\tau)$ flips the output signal of the gate $\beta'$.
It implies that  $\sem{\SeqC'[\VV_1(\SeqC',\BB,T)]}(\vec{x}_1,\cdots, \vec{x}_k)$
and $\sem{\SeqC'[\VV_2(\SeqC',\BB,T)]}(\vec{x}_1,\cdots, \vec{x}_k)$ are the same, as
the output of the gate $\beta$ is only connected to the gate $\beta'$.
Let $\VV_2(\SeqC',\BB,T)=\VV(\SeqC',\BB,T)\cup \{\EE(\alpha,\beta',\tau')\}$, where $\tau'=\tau_{bf}$ if $\tau_{bf}\in T$,
otherwise $\tau=\tau_s$ if the output signal of the gate $\beta'$ flips from $\Gzero$ to $\Gone$ due to the fault event $\EE(\alpha,\beta,\tau)$
and $\tau=\tau_r$ if the output signal of the gate $\beta'$ flips from $\Gone$ to $\Gzero$ due to the fault event $\EE(\alpha,\beta,\tau)$.
Thus,  $\sem{\SeqC'}(\vec{x}_1,\cdots, \vec{x}_k)$
and $\sem{\SeqC'[\VV_2(\SeqC',\BB,T)]}(\vec{x}_1,\cdots, \vec{x}_k)$ differ at
some clock cycle before the error flag output $o_{\tt flag}$ differs. The result immediately follows.
\end{itemize}

Moreover, if $\VV(\SeqC',\BB,T)=\emptyset$, then the output signal of the gate $\beta'$ must differ in the circuits
$\SeqC'$ and $\SeqC'[\VV_1(\SeqC',\BB,T)]$ under the same sequence of primary inputs $(\vec{x}_{1},\cdots,\vec{x}_{k})$,
otherwise $\VV_1(\SeqC',\BB,T)$ is ineffective on the circuit $\SeqC'$.
The result follows from the fact that $\VV_2(\SeqC',\BB,T)=\{\EE(\alpha,\beta',\tau'\}$.
\end{proof}

\begin{figure}[t]
\centering
\begin{lstlisting}[   %
%	basicstyle=\ttfamily, %
%	breaklines=true, %
%	keywordstyle=\bfseries\color{NavyBlue}, %
%	morekeywords={}, %
%	emph={self}, %
%    emphstyle=\bfseries\color{Rhodamine}, %
%    commentstyle=\itshape\color{black!50!white}, %
%    stringstyle=\bfseries\color{PineGreen!90!black}, %
%    columns=flexible,
%    numbersep=2em, %
%    numberstyle=\footnotesize, %
%    frame=single, %
%    framesep=1em, %
%
	language=C,
    multicols=2]
RECTANGLE S-box
Input: a,b,c,d
Output: w,x,y,z,flag
 s1 = b $\oplus$ c
 s2 = $\neg$ c
 s3 = b $\oplus$ a
 s4 = s2 $\wedge$ d
 s5 = s2 $\vee$ a
 z  = s3 $\oplus$ s4
 s6 = s5 $\oplus$ d
 s7 = s3 $\wedge$ s6
 s8 = s1 $\vee$ z
 w  = s1 $\oplus$ s7
 x  = s8 $\oplus$ s6
 y  = b $\oplus$ s6
 p1 = c $\noplus$ d  p2 = a $\vee$ c
 p3 = a $\wedge$ p1
 p4 = p2 $\nwedge$ d
 p5 = p4 $\wedge$ b
 p6 = p3 $\vee$ p5
 c1 = w $\oplus$ x  c2 = y $\oplus$ z
 c3 = c1 $\oplus$ c2
 flag = c3 $\oplus$ p6
\end{lstlisting}
\vspace{-2mm}
\caption{Pseudo-code of the illustrating example.\label{fig:examplecode}}\vspace{-3mm}
\end{figure}

\begin{figure}[t]
    \centering
    \includegraphics[width=.9\linewidth]{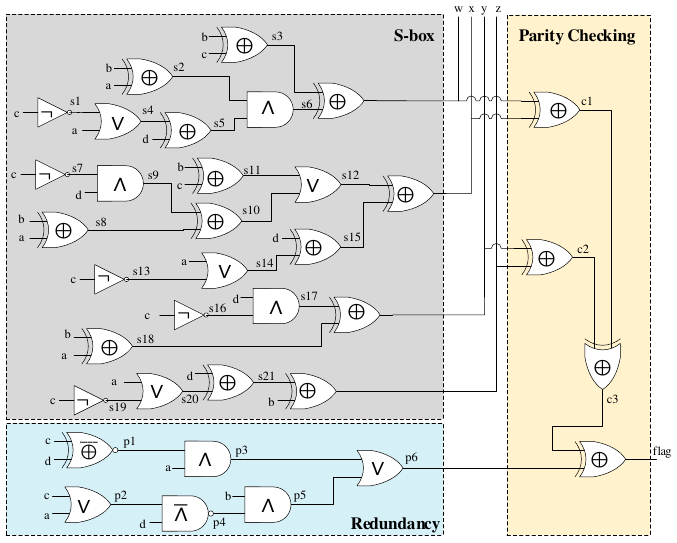} \vspace{-2mm}
    \caption{Circuit representation of the revised illustrating example.}
    \vspace{-4mm}
    \label{fig:revisedexamplecircuit}
\end{figure}

\begin{figure}[t]
\centering
\begin{lstlisting}[   %
%	basicstyle=\ttfamily, %
%	breaklines=true, %
%	keywordstyle=\bfseries\color{NavyBlue}, %
%	morekeywords={}, %
%	emph={self}, %
%    emphstyle=\bfseries\color{Rhodamine}, %
%    commentstyle=\itshape\color{black!50!white}, %
%    stringstyle=\bfseries\color{PineGreen!90!black}, %
%    columns=flexible,
%    numbersep=2em, %
%    numberstyle=\footnotesize, %
%    frame=single, %
%    framesep=1em, %
%
	language=C,
    multicols=2]
RECTANGLE S-box
Input: a,b,c,d
Output: w,x,y,z,flag
 s1 = $\neg$ c
 s2 = a $\oplus$ b
 s3 = b $\oplus$ c
 s4 = s1 $\vee$ a
 s5 = s4 $\oplus$ d
 s6 = s2 $\wedge$ s5
 s7 = $\neg$ c
 s8 = a $\oplus$ b
 s9 = s7 $\wedge$ d
 s10 = s8 $\oplus$ s9
 s11 = b $\oplus$ c
 s12 = s11 $\vee$ s12
 s13 = $\neg$ c
 s14 = a $\vee$ s13
 s15 = d $\oplus$ s14
 s16 = $\neg$ c
 s17 = d $\wedge$ s16
 s18 = a $\oplus$ b
 s19 = $\neg$ c
 s20 = a $\vee$ s19
 s21 = d $\oplus$ s20
 w = s3 $\oplus$ s6
 x = s12 $\oplus$ s15
 y = s17 $\oplus$ s18
 z = s12 $\oplus$ b
 p1 = c $\noplus$ d
 p2 = a $\vee$ c
 p3 = a $\wedge$ p1
 p4 = p2 $\nwedge$ d
 p5 = p4 $\wedge$ b
 p6 = p3 $\vee$ p5
 c1 = w $\oplus$ x
 c2 = y $\oplus$ z
 c3 = c1 $\oplus$ c2
 flag = c3 $\oplus$ p6
\end{lstlisting}
\vspace{-2mm}
\caption{Pseudo-code of the revised illustrating example.}
\vspace{-2mm}
\label{fig:revisedexamplecode}
\end{figure}

\section{Pseudo-code of the illustrating example and its Revised Version}
\label{sec:revisedexample}

The corresponding pseudo-code of the illustrating example is given in Fig.~\ref{fig:examplecode}, where the left two columns implements the function
of the S-box and the right column implements  a single-bit parity protection mechanism.

The circuit representation of the revised implementation of the RECTANGLE S-box
is shown in Fig.~\ref{fig:revisedexamplecircuit} and its pseudo-code is shown in Fig.~\ref{fig:revisedexamplecode},
following the independence property defined by~\cite{Aghaie2020ImpeccableC}.

We can observe that any fault injection on one single logic gate in the redundancy part does not change any of the outputs $\{{\tt w,x,y,z}\}$,
any fault injection on one single logic gate in the S-box part only change one of the outputs $\{{\tt w,x,y,z}\}$
and also changes the error flag output ${\tt flag}$. Thus, the revised implementation is fault-resistant
w.r.t. the blacklist $\BB$ and the fault-resistance model $\zeta(1,1,\mathcal{T},\Cc)$, where $\BB$ only contains the logic gates in the parity checking.

\section{Detailed Results of AES-R1-b1-D, AES-R1-b2-D, and CRAFT-R4-b3-D by varying the number of threads}

Detailed results of AES-R1-b1-D, AES-R1-b2-D, and CRAFT-R4-b3-D by varying the number of threads
from 1 to 12 are reported in Table~\ref{tab:multithreadsmix}.

\begin{table*}
  \centering
  \caption{Results of verifying fault-resistance of AES-R1-b1-D, AES-R1-b2-D, and CRAFT-R4-b3-D by varying the number of threads.}
\scalebox{0.9}{    \begin{tabular}{l|rrrrrrr|r}
    \toprule
     \multirow{2}{*}{\bf Name} &  \multicolumn{7}{c|}{\tool ({\bf SAT})} & \multirow{2}{*}{\bf SMT}\\  \cmidrule{2-8}
      & \textbf{1 thread} & \textbf{2 threads} & \textbf{4 threads} & \textbf{6 threads} & \textbf{8 threads} & \textbf{10 threads} & \textbf{12 threads} &   \\
    \midrule
    AES-R1-b1-$\Nn_e$1 & 1234.31  & 662.06  & 385.52  & 258.38  & 195.24  & 179.58  & \textbf{146.25}  & 4856.58  \\
    AES-R1-b1-$\Nn_e$2 & \textbf{2.34}  & 2.39  & 2.44  & 2.50  & 2.57  & 2.65  & 2.70  & 3650.83  \\
    AES-R1-b2-$\Nn_e$2 & 11028.35  & 6218.07  & 3317.16  & 2329.37  & 1850.07  & 1536.98  & \textbf{1364.48}  & 23542.57  \\
    AES-R1-b2-$\Nn_e$3 & 5.16  & \textbf{5.00}  & 5.05  & 5.21  & 5.28  & 5.41  & 5.46  & 3386.38  \\
    \midrule
    CRAFT-R4-b3-$\Nn_e$3-$\Nn_c$1 & 453.57  & 264.05  & 136.66  & 94.12  & 62.07  & 55.14  & \textbf{45.52}  & 1188.03  \\
    CRAFT-R4-b3-$\Nn_e$4-$\Nn_c$1 & 0.73  & \textbf{0.71}  & 0.74  & 0.76  & 0.77  & 0.80  & 0.81  & 8.25  \\
    CRAFT-R4-b3-$\Nn_e$3-$\Nn_c$2 & 998.18  & 524.15  & 295.33  & 206.91  & 155.75  & 136.38  & \textbf{127.18}  & 978.78  \\
    CRAFT-R4-b3-$\Nn_e$3-$\Nn_c$3 & 577.90  & 290.44  & 150.73  & 102.40  & 82.08  & 66.91  & \textbf{55.21}  & 900.28  \\
    CRAFT-R4-b3-$\Nn_e$3-$\Nn_c$4 & 398.52  & 217.15  & 120.24  & 76.40  & 71.15  & 54.40  & \textbf{45.34}  & 916.13  \\
    CRAFT-R4-b3-$\Nn_e$3-$\Nn_c$5 & 396.56  & 208.42  & 103.70  & 75.11  & 55.16  & 46.08  & \textbf{37.08}  & 922.49  \\
    \bottomrule
    \end{tabular}}%
    \label{tab:multithreadsmix}
\end{table*}%

\section{Results of Fault-resistance Verification with only the Fault Type $\tau_{bf}$}
We also evaluate \tool for verifying fault-resistance with only the fault type $\tau_{bf}$.

 \begin{table*}
    \centering\setlength\tabcolsep{4pt}
  \caption{Results of fault-resistance verification with only the fault type $\tau_{bf}$.}\vspace{-3mm}
  \label{tab:fault-resistancebf}
\scalebox{0.9}{
    \begin{tabular}{l|c|rrrrr|r|r|c|c } \toprule
      \multirow{2}{*}{\bf Name}   & {\bf Fault-resistance} & \multicolumn{5}{c|}{\tool ({\bf SAT})} & \textbf{SMT} & \textbf{FIVER} & \multirow{2}{*}{\bf Result} &{\bf Desired} \\ \cline{3-9}
         & {\bf model} & \textbf{\#Var} & \textbf{\# Clause} & \textbf{2CNF} & \textbf{Solving} & \textbf{Total} & \textbf{Time} & \textbf{Time} &  &{\bf Result}\\ \midrule
        AES-R1-b1-D & $\zeta (1,1, \tau_{bf},\CR)$ & 34822 & 246273 & 1.87  & 159.44  & 161.31  & 4728.67  & \textbf{24.50}  & \ding{51}  & \ding{51} \\
        AES-R1-b1-D & $\zeta (2,1,\tau_{bf},\CR)$ & 46570 & 181812 & 2.33  & 0.41  & \textbf{2.74}  & 3192.99 & 10154.53  & \myding{55}& \myding{55}\\
        AES-R1-b2-D & $\zeta (2,1,\tau_{bf},\CR)$ & 56196 & 243722 & 4.50  & 1397.42  & \textbf{1401.92}  & 24397.81  & timeout & \ding{51} & \ding{51}  \\
        AES-R1-b2-D & $\zeta (3,1,\tau_{bf},\CR)$ & 60291 & 268197 & 4.44  & 0.66  & \textbf{5.10}  & 5254.88 & timeout & \myding{55}& \myding{55}\\ \midrule
        CRAFT-R1-b1-C & $\zeta (1,1,\tau_{bf},\CR)$ & 4326 & 26659 & 0.09  & 0.64  & 0.73  & 5.87  & \textbf{0.10}  & \ding{51}  & \ding{51} \\
        CRAFT-R1-b1-C & $\zeta (2,1,\tau_{bf},\CR)$ & 5167 & 24760 & 0.09  & 0.06  & \textbf{0.15}  & 0.16 & 1.03  & \myding{55}& \myding{55}\\
        CRAFT-R1-b2-C & $\zeta (2,1,\tau_{bf},\CR)$ & 35997 & 188158 & 2.11  & 142.04  & \textbf{144.15}  & 2586.82  & 175.84  & \ding{51}  & \ding{51} \\
        CRAFT-R1-b2-C & $\zeta (3,1,\tau_{bf},\CR)$ & 38195 & 212732 & 2.24  & 0.56  & \textbf{2.80}  & 9.79 & 958.40  & \myding{55}& \myding{55}\\
        CRAFT-R1-b1-D & $\zeta (1,1,\tau_{bf},\CR)$ & 1347 & 7890 & 0.02  & 0.16  & 0.18  & 0.36  & \textbf{0.03}  & \ding{51}  & \ding{51} \\
        CRAFT-R1-b1-D & $\zeta (2,1,\tau_{bf},\CR)$ & 1659 & 7471 & 0.03  & 0.02  & 0.05  & \textbf{0.01}  & 0.06  & \myding{55}\\
        CRAFT-R1-b2-D & $\zeta (2,1,\tau_{bf},\CR)$ & 2165 & 10140 & 0.03  & 0.31  & 0.34  & 0.61  & \textbf{0.17}  & \ding{51}  & \ding{51} \\
        CRAFT-R1-b2-D & $\zeta (3,1,\tau_{bf},\CR)$ & 2347 & 11418 & 0.04  & 0.04  & 0.08  & \textbf{0.01}  & 8.64  & \myding{55}& \myding{55}\\
        CRAFT-R1-b3-D & $\zeta (3,1,\tau_{bf},\CR)$ & 2572 & 14530 & 0.04  & 0.64  & \textbf{0.68}  & 5.12  & 310.40  & \ding{51} & \ding{51}  \\
        CRAFT-R1-b3-D & $\zeta (4,1,\tau_{bf},\CR)$ & 2730 & 15916 & 0.04  & 0.05  & \textbf{0.09}  & 0.58 & 1724.35 & \myding{55}& \myding{55}\\  \midrule
        CRAFT-R2-b1-C & $\zeta (1,1,\tau_{bf},\CR)$ & 8008 & 49988 & 0.18  & 13.44  & 13.62  & 91.76  & \textbf{2.09}  & \ding{51} & \ding{51}  \\
        CRAFT-R2-b1-C & $\zeta (2,1,\tau_{bf},\CR)$ & 9579 & 46410 & 0.20  & 0.13  & \textbf{0.33}  & 0.42 & 41.27  & \myding{55}& \myding{55}\\
        CRAFT-R2-b2-C & $\zeta (2,1,\tau_{bf},\CR)$ & 70834 & 361687 & 5.50  & 1891.50	& 1897.00  & 12779.36 & \textbf{1770.58} & \ding{51}  & \ding{51} \\
        CRAFT-R2-b2-C & $\zeta (3,1,\tau_{bf},\CR)$ & 75079 & 407564 & 5.60  & 2.20  & \textbf{7.80}  & 2464.15 & 3317.16 & \myding{55}& \myding{55}\\
        CRAFT-R2-b1-D & $\zeta (1,1,\tau_{bf},\CR)$ & 2595 & 15155 & 0.04  & 0.91  & 0.95  & 1.89  & \textbf{0.64}  & \ding{51} & \ding{51} \\
        CRAFT-R2-b1-D & $\zeta (2,1,\tau_{bf},\CR)$ & 3195 & 14214 & 0.03  & 0.04  & 0.07  & \textbf{0.04}  & 1.05  & \myding{55}& \myding{55}\\
        CRAFT-R2-b1-D & $\zeta (1,2,\tau_{bf},\CR)$ & 2595 & 15155 & 0.05  & 0.85  & \textbf{0.90}  & 3.63  & 124.40  & \ding{51}  & \ding{51} \\
        CRAFT-R2-b1-D & $\zeta (1,3,\tau_{bf},\CR)$ & 2594 & 15153 & 0.06  & 0.62  & \textbf{0.68}  & 3.53  & 25815.91  & \ding{51} & \ding{51}  \\
        CRAFT-R2-b2-D & $\zeta (2,1,\tau_{bf},\CR)$ & 4015 & 19410 & 0.09  & 0.94  & \textbf{1.03}  & 6.03  & 6.50  & \ding{51}  & \ding{51} \\
        CRAFT-R2-b2-D & $\zeta (3,1,\tau_{bf},\CR)$ & 4324 & 21975 & 0.05  & 0.06  & 0.11  & \textbf{0.05}  & 89.51  & \myding{55}& \myding{55}\\
        CRAFT-R2-b2-D & $\zeta (2,2,\tau_{bf},\CR)$ & 4015 & 19410 & 0.14  & 1.02  & \textbf{1.16}  & 8.63  & timeout & \ding{51}  & \ding{51} \\   \midrule
        CRAFT-R3-b3-D & $\zeta (3,1,\tau_{bf},\CR)$ & 7066 & 38313 & 0.17  & 13.11  & \textbf{13.28}  & 241.14  & timeout & \ding{51}  & \ding{51} \\
        CRAFT-R3-b3-D & $\zeta (4,1,\tau_{bf},\CR)$ & 7462 & 41709 & 0.17  & 0.11  & \textbf{0.28}  & 3.19  & timeout & \myding{55}& \myding{55}\\
        CRAFT-R3-b3-D & $\zeta (3,2,\tau_{bf},\CR)$ & 7068 & 38332 & 0.17  & 15.08  & \textbf{15.25}  & 243.42  & timeout & \ding{51}  & \ding{51} \\
        CRAFT-R3-b3-D & $\zeta (3,3,\tau_{bf},\CR)$ & 7066 & 38313 & 0.18  & 16.36  & \textbf{16.54}  & 242.72  & timeout & \ding{51}  & \ding{51} \\
        CRAFT-R3-b3-D & $\zeta (3,4,\tau_{bf},\CR)$ & 7065 & 38309 & 0.15  & 14.32  & \textbf{14.47}  & 241.82 & timeout & \ding{51}  & \ding{51} \\  \midrule
        CRAFT-R4-b3-D & $\zeta (3,1,\tau_{bf},\CR)$ & 9209 & 50064 & 0.60  & 23.10  & \textbf{23.70}  & 717.94  & timeout & \ding{51}  & \ding{51} \\
        CRAFT-R4-b3-D & $\zeta (4,1,\tau_{bf},\CR)$ & 9732 & 54521 & 0.62  & 0.16  & \textbf{0.78}  & 6.87  & timeout & \myding{55}& \myding{55}\\
        CRAFT-R4-b3-D & $\zeta (3,2,\tau_{bf},\CR)$ & 9211 & 50095 & 0.59  & 34.93  & \textbf{35.52}  & 667.81  & timeout & \ding{51}  & \ding{51} \\
        CRAFT-R4-b3-D & $\zeta (3,3,\tau_{bf},\CR)$ & 9211 & 50095 & 0.56  & 26.07  & \textbf{26.63}  & 671.85  & timeout & \ding{51}  & \ding{51} \\
        CRAFT-R4-b3-D & $\zeta (3,4,\tau_{bf},\CR)$ & 9209 & 50064 & 0.58  & 25.43  & \textbf{26.01}  & 688.15  & timeout & \ding{51}  & \ding{51} \\
        CRAFT-R4-b3-D & $\zeta (3,5,\tau_{bf},\CR)$ & 9208 & 50057 & 0.57  & 25.95  & \textbf{26.52}  & 691.08 & timeout & \ding{51}  & \ding{51} \\   \midrule
        LED64-R1-b1-D & $\zeta (1,1,\tau_{bf},\CR)$ & 1866 & 10623 & 0.06  & 0.46  & 0.52  & 11.63  & \textbf{0.07}  & \ding{51}  & \ding{51} \\
        LED64-R1-b1-D & $\zeta (2,1,\tau_{bf},\CR)$ & 2231 & 9401 & 0.05  & 0.02  & 0.07  & \textbf{0.02}  & 0.74  & \myding{55}& \myding{55}\\
        LED64-R1-b2-D & $\zeta (2,1,\tau_{bf},\CR)$ & 2628 & 12473 & 0.09  & 1.54  & 1.63  & 12.01  & \textbf{1.40}  & \ding{51}  & \ding{51} \\
        LED64-R1-b2-D & $\zeta (3,1,\tau_{bf},\CR)$ & 2755 & 13268 & 0.07  & 0.03  & 0.10  & \textbf{0.03}  & 32.71  & \myding{55}& \myding{55}\\  \midrule
        LED64-R2-b1-D & $\zeta (1,1,\tau_{bf},\CR)$ & 3594 & 21569 & 0.12  & 1.68  & \textbf{1.80}  & 54.22  & timeout & \ding{51}  & \ding{51} \\
        LED64-R2-b1-D & $\zeta (2,1,\tau_{bf},\CR)$ & 4162 & 20409 & 0.09  & 0.07  & \textbf{0.16}  & 1.44  & timeout & \myding{55}& \myding{55}\\
        LED64-R2-b2-D & $\zeta (2,1,\tau_{bf},\CR)$ & 5116 & 26420 & 0.12  & 3.21  & \textbf{3.33}  & 266.12  & timeout & \ding{51} & \ding{51}  \\
        LED64-R2-b2-D & $\zeta (3,1,\tau_{bf},\CR)$ & 5370 & 29218 & 0.14  & 0.08  & \textbf{0.22}  & 2.21  & timeout & \myding{55}& \myding{55}\\  \midrule
        LED64-R3-b1-D & $\zeta (1,1,\tau_{bf},\CR)$ & 5249 & 31588 & 0.13  & 4.28  & \textbf{4.41}  & 263.93  & timeout & \ding{51}  & \ding{51} \\
        LED64-R3-b1-D & $\zeta (2,1,\tau_{bf},\CR)$ & 6103 & 29921 & 0.13  & 0.11  & \textbf{0.24}  & 2.96 & timeout & \myding{55}& \myding{55}\\
        LED64-R3-b2-D & $\zeta (2,1,\tau_{bf},\CR)$ & 7735 & 38318 & 0.12  & 7.79  & \textbf{7.91}  & 394.05  & timeout & \ding{51} & \ding{51}  \\
        LED64-R3-b2-D & $\zeta (3,1,\tau_{bf},\CR)$ & 8116 & 41939 & 0.22  & 0.15  & \textbf{0.37}  & 3.06 & timeout & \myding{55}& \myding{55}\\ \bottomrule
    \end{tabular}}\vspace{-3mm}
\end{table*}

\subsection{RQ1: Efficiency and Effectiveness of \tool for Fault-resistance Verification}
We still compare \tool with (1) the state-of-the-art verifier FIVER and (2) an SMT-based approach which directly
checks the constraints generated by our encoding method without translating to Boolean formulas.

The results are reported in Table~\ref{tab:fault-resistancebf}, where both the SAT solver Glucose and the BDD-based verifier FIVER are run with 8 threads
while the SMT solver bitwuzla is run with a single thread.
Columns (2CNF) and (Solving) give the execution time of building and solving Boolean formulas  in seconds, respectively.
Columns (Total) and (Time) give the total execution time in seconds.
Mark \ding{51} (resp. \myding{55}) indicates that the protected circuit $\SeqC'$ is fault-resistant (resp. not fault-resistant).

Overall, both \tool (i.e., the SAT-based approach) and the SMT-based approach solved all the verification tasks, while FIVER runs out of time on 22 verification tasks within the time limit (24 hours per task). The SAT/SMT-based approach become more and more efficient than FIVER
with the increase of round numbers (i.e., R$i$) and maximal number of protected faulty bits (i.e., b$j$).
\tool is significantly more efficient than the SMT-based approach on relatively larger benchmarks (e.g., AES-R1-b1, AES-R1-b2, CRAFT-R1-b2-C, CRAFT-R2-b2-C, CRAFT-R3-b3-D, CRAFT-R4-b3-D, LED64-R2-b1-D, LED64-R2-b2-D, LED64-R3-b1-D and LED64-R3-b2-D) while they are comparable on smaller benchmarks.

Interestingly, we found that
(i)  implementations with correction-based countermeasures are more difficult to prove than that with detection-based countermeasures (e.g., CRAFT-R$i$-b$j$-C vs. CRAFT-R$i$-b$j$-D, for $i=1,2$ and $j=1,2$), because implementing correction-based countermeasures require more gates;
(ii) \tool is more efficient at disproving fault-resistance than proving fault-resistance, because UNSAT instances are often more difficult to prove than SAT instances in CDCL SAT solvers.
(iii) \tool often scales very well with increase of the round numbers (i.e., R$i$ for $i=1,2,3,4$),
maximal number of protected faulty bits (i.e., b$j$ for $j=1,2,3$), maximum number of fault events per clock cycle (i.e., $\Nn_e$)
and the maximum number of clock cycles in which fault events can occur  (i.e., $\Nn_c$),
but FIVER has very limited scalability.

\begin{figure}[t]
    \centering
    \subfigure[AES-R1-b1-D and AES-R1-b2-D]{
    \label{fig:AESthreads}
    \includegraphics[scale=0.45]{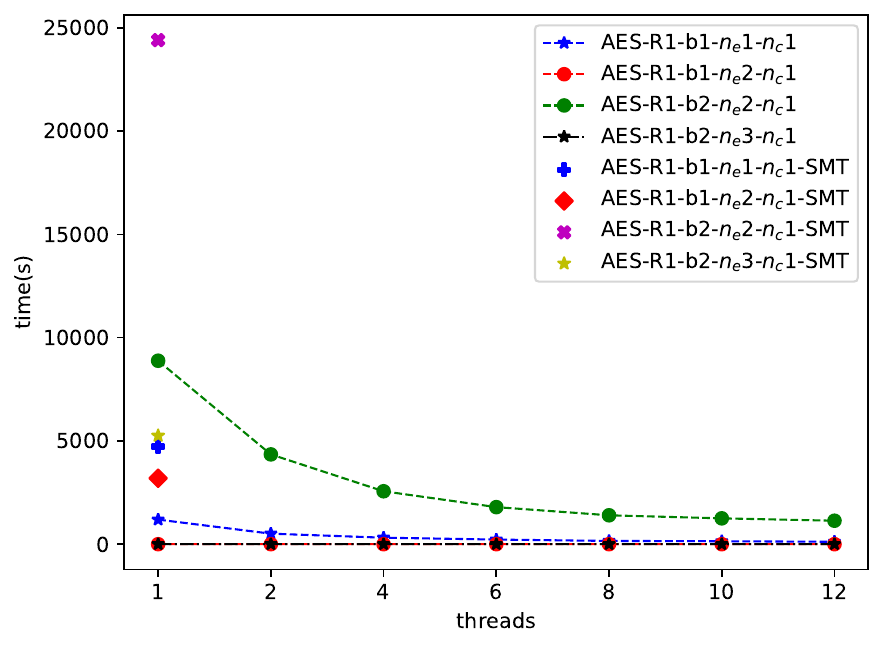}}
    \subfigure[CRAFT-R4-b3-D]{
    \label{fig:CRAFTthreads}
    \includegraphics[scale=0.45]{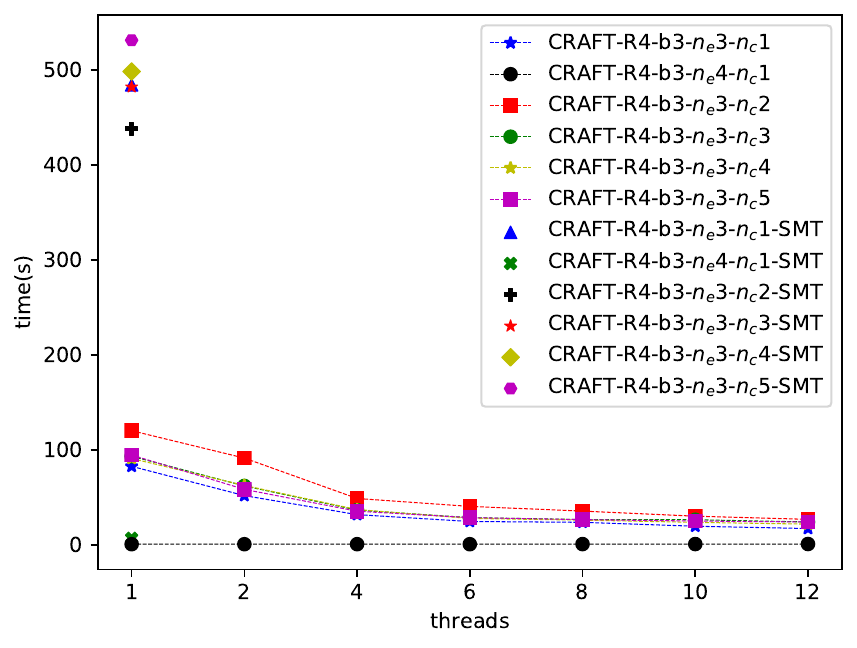}}\vspace{-4mm}
    \caption{Comparison of verification times by varying the number of threads with only the fault type $\tau_{bf}$}
\end{figure}

\begin{table*}
  \centering\setlength\tabcolsep{6pt}
  \caption{Results of verifying fault-resistance of AES-R1-b1-D, AES-R1-b2-D, and CRAFT-R4-b3-D with only the fault type $\tau_{bf}$ by varying the number of threads.}\vspace{-3mm}
\scalebox{0.9}{
 \begin{tabular}{l|rrrrrrr|r}
    \toprule
     \multirow{2}{*}{\bf Name} &  \multicolumn{7}{c|}{\tool ({\bf SAT})} & \multirow{2}{*}{\bf SMT}\\  \cmidrule{2-8}
      & \textbf{1 thread} & \textbf{2 threads} & \textbf{4 threads} & \textbf{6 threads} & \textbf{8 threads} & \textbf{10 threads} & \textbf{12 threads} &   \\
    \midrule
    AES-R1-b1-$\Nn_e$1 & 1192.68 & 513.15 & 317.66 & 226.61 & 161.31 & 145.69 & \textbf{119.36} & 4728.67 \\
    AES-R1-b1-$\Nn_e$2 & 2.71  & \textbf{2.70}   & 2.74  & 2.84  & 2.74  & 2.91  & 3.02  & 3192.99 \\
    AES-R1-b2-$\Nn_e$2 & 8882.96 & 4349.25 & 2564.76 & 1796.85 & 1401.92 & 1253.49 & \textbf{1139.42} & 24397.81 \\
    AES-R1-b2-$\Nn_e$3 & \textbf{6.59}  & 6.78  & 6.67  & 6.77  & 6.67  & 6.99  & 7.08  & 5254.88 \\ \midrule
    CRAFT-R4-b3-$\Nn_e$3-$\Nn_c$1 & 82.61 & 51.79 & 31.79 & 24.56 & 23.70 & 19.56 & \textbf{17.15} & 484.17 \\
    CRAFT-R4-b3-$\Nn_e$4-$\Nn_c$1 & \textbf{0.67}  & \textbf{0.67}  & 0.69  & \textbf{0.67}  & 0.78 & 0.81  & 0.86  & 6.87 \\
    CRAFT-R4-b3-$\Nn_e$3-$\Nn_c$2 & 120.11 & 91.31 & 48.79 & 40.44 & 35.52 & 30.15 & \textbf{26.83} & 437.81 \\
    CRAFT-R4-b3-$\Nn_e$3-$\Nn_c$3 & 93.24 & 62.00    & 36.55 & 28.61 & 26.63 & 26.40  & \textbf{24.06} &481.85 \\
    CRAFT-R4-b3-$\Nn_e$3-$\Nn_c$4 & 90.88 & 62.49 & 37.16 & 27.81 & 26.01 & 23.66 & \textbf{21.79} & 498.15 \\
    CRAFT-R4-b3-$\Nn_e$3-$\Nn_c$5 & 94.53 & 58.37 & 35.38 & 28.61 & 26.52 & 25.07 & \textbf{24.04} & 531.08 \\
    \bottomrule
    \end{tabular}}%
    \label{tab:multithreads}
\end{table*}%

To understand the effect of the number of threads, we evaluate \tool on the benchmarks AES-R1-b1-D, AES-R1-b2-D, and CRAFT-R4-b3-D
by varying the number of threads from 1 to 12. The results are depicted in Fig.~\ref{fig:AESthreads}
and Fig.~\ref{fig:CRAFTthreads}, respectively, where $\Nn_ei$ and $\Nn_cj$ denote the fault-resistance mode $\zeta(i,j,\tau_{bf},\CR)$.
Detailed results are reported in Table~\ref{tab:multithreads}.
We can observe that \tool always outperforms the SMT-based approach.
On the fault-resistant benchmarks (i.e., curves with b$j$-$\Nn_e k$ such that $j\geq k$), \tool becomes more and more efficient
while the improvement becomes less and less, with the increase of the number of threads.
On the non-fault-resistant benchmarks (i.e., curves with b$j$-$\Nn_e k$ such that $j<k$), multi-threading does not improve performance
and instead slightly worsens performance, because they are easy to be disproved and thread scheduling causes overhead.

\begin{table*}
\centering\setlength\tabcolsep{6pt}
  \caption{Results of fault-resistance verification using \tool with/without vulnerable gate reduction using the fault type $\tau_{bf}$.}\vspace{-3mm}
  \label{tab:faultgatereductionbf}
\scalebox{0.9}{
    \begin{tabular}{l|c| rrrr| rrrr|c|c}    \toprule
  \multirow{2}{*}{\bf Name} & \textbf{Fault-resistance} & \multicolumn{4}{c|}{\bf Without vulnerable gate reduction}  & \multicolumn{4}{c|}{\bf With vulnerable gate reduction}  &  \multirow{2}{*}{\bf Result} & {\bf Desired}\\ \cline{3-10}
                             & \textbf{model} &  \textbf{\#Var} & \textbf{\#Clause} & \textbf{\#Gate}  & \textbf{Time} & \textbf{\#Var} & \textbf{\#Clause} & \textbf{\#Gate} & \textbf{Time} & & {\bf Result}\\  \midrule
        AES-R1-b1-D   & $\zeta (1,1,\tau_{bf},\CR)$ & 77675  & 577979 & 24432  & 972.45  & 34822 & 246273 & {\bf 7920}   & {\bf 161.31} & \ding{51} & \ding{51} \\
        AES-R1-b1-D   & $\zeta (2,1,\tau_{bf},\CR)$ & 100592 & 417708 & 24432  & 3.52    & 46570 & 181812 & {\bf 7920}   & {\bf 2.74} & \myding{55} & \myding{55}  \\
        AES-R1-b2-D   & $\zeta (2,1,\tau_{bf},\CR)$ & 149413 & 466954 & 33104  & timeout & 56196 & 243722 & {\bf 10640}  & {\bf 1401.92} & \ding{51}& \ding{51}  \\
        AES-R1-b2-D   & $\zeta (3,1,\tau_{bf},\CR)$ & 165796 & 517451 & 33104  & 6.09    & 60291 & 268197 & {\bf 10640}  & {\bf 5.10} & \myding{55}  & \myding{55} \\    \midrule
        CRAFT-R1-b1-C & $\zeta (1,1,\tau_{bf},\CR)$ & 9773   & 70431  & 2948   & 3.71    & 4326  & 26659  & {\bf 760}    & {\bf 0.73} & \ding{51} & \ding{51} \\
        CRAFT-R1-b1-C & $\zeta (2,1,\tau_{bf},\CR)$ & 12422  & 61020  & 2948   & 0.28    & 5167  & 24760  & {\bf 760}    & {\bf 0.15} & \myding{55}  & \myding{55} \\
        CRAFT-R1-b2-C & $\zeta (2,1,\tau_{bf},\CR)$ & 88296  & 415886 & 19636  & 2263.07 & 35997 & 188158 & {\bf 5672}   & {\bf 144.15} & \ding{51} & \ding{51} \\
        CRAFT-R1-b2-C & $\zeta (3,1,\tau_{bf},\CR)$ & 96746  & 500084 & 19636  & 4.10    & 38195 & 212732 & {\bf 5672}   & {\bf 2.80} & \myding{55}  & \myding{55} \\
        CRAFT-R1-b1-D & $\zeta (1,1,\tau_{bf},\CR)$ & 2657   & 17579  & 766    & 0.26    & 1347  & 7890   & {\bf 274}    & {\bf 0.18} & \ding{51} & \ding{51} \\
        CRAFT-R1-b1-D & $\zeta (2,1,\tau_{bf},\CR)$ & 3493   & 15509  & 766    & 0.08    & 1659  & 7471   & {\bf 274}    & {\bf 0.05} & \myding{55}  & \myding{55} \\
        CRAFT-R1-b2-D & $\zeta (2,1,\tau_{bf},\CR)$ & 4716   & 22755  & 1139   & 1.32    & 2165  & 10140  & {\bf 376}    & {\bf 0.34} & \ding{51} & \ding{51} \\
        CRAFT-R1-b2-D & $\zeta (3,1,\tau_{bf},\CR)$ & 5214   & 27519  & 1139   & 0.12    & 2347  & 11418  & {\bf 376}    & {\bf 0.08} & \myding{55} & \myding{55}  \\
        CRAFT-R1-b3-D & $\zeta (3,1,\tau_{bf},\CR)$ & 5990   & 31598  & 1296   & 10.24   & 2572  & 14530  & {\bf 448}    & {\bf 0.68} & \ding{51} & \ding{51} \\
        CRAFT-R1-b3-D & $\zeta (4,1,\tau_{bf},\CR)$ & 6468   & 35656  & 1296   & 0.18    & 2730  & 15916  & {\bf 448}    & {\bf 0.09} & \myding{55} & \myding{55}  \\   \midrule
        CRAFT-R2-b1-C & $\zeta (1,1,\tau_{bf},\CR)$ & 18430  & 134017 & 5656   & 71.00   & 8008  & 49988  & {\bf 1440}   & {\bf 13.62} & \ding{51} & \ding{51} \\
        CRAFT-R2-b1-C & $\zeta (2,1,\tau_{bf},\CR)$ & 23948  & 115187 & 5656   & 0.76    & 9579  & 46410  & {\bf 1440}   & {\bf 0.33} & \myding{55}  & \myding{55} \\
        CRAFT-R2-b2-C & $\zeta (2,1,\tau_{bf},\CR)$ & 172699 & 818270 & 38872  & timeout & 70834 & 361687 & {\bf 11200}  & {\bf 1897.00} & \ding{51} & \ding{51} \\
        CRAFT-R2-b2-C & $\zeta (3,1,\tau_{bf},\CR)$ & 189340 & 988323 & 38872  & 10.88   & 75079 & 407564 & {\bf 11200}  & {\bf 7.80} & \myding{55}  & \myding{55} \\
        CRAFT-R2-b1-D & $\zeta (1,1,\tau_{bf},\CR)$ & 5169   & 34088  & 1480   & 3.14    & 2595  & 15155  & {\bf 508}    & {\bf 0.95} & \ding{51} & \ding{51} \\
        CRAFT-R2-b1-D & $\zeta (2,1,\tau_{bf},\CR)$ & 7058   & 29782  & 1480   & 0.16    & 3195  & 14214  & {\bf 508}    & {\bf 0.07} & \myding{55} & \myding{55}  \\
        CRAFT-R2-b1-D & $\zeta (1,2,\tau_{bf},\CR)$ & 5169   & 34088  & 1480   & 4.32    & 2595  & 15155  & {\bf 508}    & {\bf 0.90} & \ding{51} & \ding{51} \\
        CRAFT-R2-b1-D & $\zeta (1,3,\tau_{bf},\CR)$ & 5168   & 34086  & 1480   & 3.59    & 2594  & 15153  & {\bf 508}    & {\bf 0.68} & \ding{51} & \ding{51} \\
        CRAFT-R2-b2-D & $\zeta (2,1,\tau_{bf},\CR)$ & 9294   & 43390  & 2180   & 58.66   & 4015  & 19410  & {\bf 672}    & {\bf 1.03} & \ding{51} & \ding{51} \\
        CRAFT-R2-b2-D & $\zeta (3,1,\tau_{bf},\CR)$ & 10303  & 51899  & 2180   & 0.25    & 4324  & 21975  & {\bf 672}    & {\bf 0.11} & \myding{55} & \myding{55}  \\
        CRAFT-R2-b2-D & $\zeta (2,2,\tau_{bf},\CR)$ & 9294   & 43390  & 2180   & 33.45   & 4015  & 19410  & {\bf 672}    & {\bf 1.16} & \ding{51} & \ding{51} \\   \midrule
        CRAFT-R3-b3-D & $\zeta (3,1,\tau_{bf},\CR)$ & 17284  & 92557  & 3776   & 1274.24 & 7066  & 38313  & {\bf 1104}   & {\bf 13.28} & \ding{51} & \ding{51} \\
        CRAFT-R3-b3-D & $\zeta (4,1,\tau_{bf},\CR)$ & 18656  & 104353 & 3776   & 0.43    & 7462  & 41709  & {\bf 1104}   & {\bf 0.28} & \myding{55} & \myding{55}  \\
        CRAFT-R3-b3-D & $\zeta (3,2,\tau_{bf},\CR)$ & 17286  & 92576  & 3776   & 2323.82 & 7068  & 38332  & {\bf 1104}   & {\bf 15.25} & \ding{51} & \ding{51} \\
        CRAFT-R3-b3-D & $\zeta (3,3,\tau_{bf},\CR)$ & 17284  & 92557  & 3776   & 1615.21 & 7066  & 38313  & {\bf 1104}   & {\bf 16.54} & \ding{51} & \ding{51} \\
        CRAFT-R3-b3-D & $\zeta (3,4,\tau_{bf},\CR)$ & 17283  & 92553  & 3776   & 1461.67 & 7065  & 38309  & {\bf 1104}   & {\bf 14.47} & \ding{51} & \ding{51} \\   \midrule
        CRAFT-R4-b3-D & $\zeta (3,1,\tau_{bf},\CR)$ & 22681  & 122758 & 4992   & 2152.98 & 9209  & 50064  & {\bf 1440}   & {\bf 23.70} & \ding{51} & \ding{51} \\
        CRAFT-R4-b3-D & $\zeta (4,1,\tau_{bf},\CR)$ & 24500  & 138447 & 4992   & 1.08    & 9732  & 54521  & {\bf 1440}   & {\bf 0.78} & \myding{55} & \myding{55}  \\
        CRAFT-R4-b3-D & $\zeta (3,2,\tau_{bf},\CR)$ & 22683  & 122789 & 4992   & 8174.81 & 9211  & 50095  & {\bf 1440}   & {\bf 35.52} & \ding{51} & \ding{51} \\
        CRAFT-R4-b3-D & $\zeta (3,3,\tau_{bf},\CR)$ & 22683  & 122789 & 4992   & 4473.37 & 9211  & 50095  & {\bf 1440}   & {\bf 26.63} & \ding{51} & \ding{51} \\
        CRAFT-R4-b3-D & $\zeta (3,4,\tau_{bf},\CR)$ & 22681  & 122758 & 4992   & 3864.15 & 9209  & 50064  & {\bf 1440}   & {\bf 26.01} & \ding{51} & \ding{51} \\
        CRAFT-R4-b3-D & $\zeta (3,5,\tau_{bf},\CR)$ & 22680  & 122751 & 4992   & 3626.20 & 9208  & 50057  & {\bf 1440}   & {\bf 26.52} & \ding{51} & \ding{51} \\   \midrule
        LED64-R1-b1-D & $\zeta (1,1,\tau_{bf},\CR)$ & 3886   & 29433  & 1312   & 3.87    & 1866  & 10623  & {\bf 240}    & {\bf 0.52} & \ding{51} & \ding{51} \\
        LED64-R1-b1-D & $\zeta (2,1,\tau_{bf},\CR)$ & 5329   & 21857  & 1312   & 0.10    & 2231  & 9401   & {\bf 240}    & {\bf 0.07} & \myding{55} & \myding{55}  \\
        LED64-R1-b2-D & $\zeta (2,1,\tau_{bf},\CR)$ & 8128   & 27465  & 1888   & 27.49   & 2628  & 12473  & {\bf 336}    & {\bf 1.63} & \ding{51} & \ding{51} \\
        LED64-R1-b2-D & $\zeta (3,1,\tau_{bf},\CR)$ & 8991   & 30212  & 1888   & 0.20    & 2755  & 13268  & {\bf 336}    & {\bf 0.10} & \myding{55} & \myding{55}  \\   \midrule
        LED64-R2-b1-D & $\zeta (1,1,\tau_{bf},\CR)$ & 8139   & 61143  & 2496   & 26.85   & 3594  & 21569  & {\bf 480}    & {\bf 1.80} & \ding{51} & \ding{51} \\
        LED64-R2-b1-D & $\zeta (2,1,\tau_{bf},\CR)$ & 10202  & 53817  & 2496   & 0.29    & 4162  & 20409  & {\bf 480}    & {\bf 0.16} & \myding{55} & \myding{55}  \\
        LED64-R2-b2-D & $\zeta (2,1,\tau_{bf},\CR)$ & 15630  & 69468  & 3536   & 279.56  & 5116  & 26420  & {\bf 672}    & {\bf 3.33} & \ding{51} & \ding{51} \\
        LED64-R2-b2-D & $\zeta (3,1,\tau_{bf},\CR)$ & 17116  & 80426  & 3536   & 0.57    & 5370  & 29218  & {\bf 672}    & {\bf 0.22} & \myding{55}  & \myding{55} \\   \midrule
        LED64-R3-b1-D & $\zeta (1,1,\tau_{bf},\CR)$ & 11963  & 90062  & 3680   & 63.25   & 5249  & 31588  & {\bf 720}    & {\bf 4.41} & \ding{51} & \ding{51} \\
        LED64-R3-b1-D & $\zeta (2,1,\tau_{bf},\CR)$ & 15464  & 77778  & 3680   & 0.45    & 6103  & 29921  & {\bf 720}    & {\bf 0.24} & \myding{55} & \myding{55}  \\
        LED64-R3-b2-D & $\zeta (2,1,\tau_{bf},\CR)$ & 22448  & 103306 & 5184   & 453.41  & 7735  & 38318  & {\bf 1008}   & {\bf 7.91} & \ding{51}& \ding{51}  \\
        LED64-R3-b2-D & $\zeta (3,1,\tau_{bf},\CR)$ & 24493  & 121295 & 5184   & 0.69    & 8116  & 41939  & {\bf 1008}   & {\bf 0.37} & \myding{55}  & \myding{55} \\   \bottomrule
    \end{tabular}}
\end{table*}

\subsection{RQ2: Effectiveness of the Vulnerable Gate Reduction}
We evaluate \tool with/without the vulnerable gate reduction using 8 threads for SAT solving,
considering only the fault type $\tau_{bf}$.

The results are reported in Table~\ref{tab:faultgatereductionbf},
where columns (\#Gate) give the number of vulnerable gates that should be considered when verifying
fault-resistance. We can observe that our vulnerable gate reduction is able to significantly reduce the number of vulnerable gates that should be considered when verifying
fault-resistance (more than 72\% reduction rate on average), consequently,  significantly reducing the size of the resulting Boolean formulas and
accelerate fault-resistance verification, no matter the adopted countermeasure, fault-resistance model,
fault-resistance and size of the benchmarks.

\end{document}